\documentclass[3p,procedia]{elsarticle}

\journal{Journal of Information Processing and Management}

\usepackage{hyperref}
\usepackage{url}
\usepackage[neverdecrease]{paralist}
\usepackage[caption=false]{subfig}
\usepackage{graphicx}
\usepackage{booktabs}
\usepackage{tabularx}
\usepackage{xspace}
\usepackage{multirow}
\usepackage{amsmath}
\usepackage{calc}
\usepackage[ruled,linesnumbered]{algorithm2e}
\usepackage{comment}
\usepackage{natbib}
\usepackage{color,soul}

\usepackage{amssymb}
\usepackage{pifont}
\usepackage{bm}
\usepackage{colortbl}
\usepackage{pgfplots}
\usepackage{pgfplotstable}

\usepackage[inline,shortlabels]{enumitem}

\newcommand{\ptag}[1]{{\sffamily #1}}%
\newcommand{\term}[1] {\emph{\textbf{#1}}}

\makeatletter
\def\BState{\State\hskip-\ALG@thistlm}
\makeatother

\makeatletter
\newcommand{\ssymbol}[1]{^{\@fnsymbol{#1}}}
\makeatother

\usepackage{array}

\makeatletter
\newlength\mylena
\newlength\mylenb
\newcommand\mystrut[1][2]{%
    \setlength\mylena{#1\ht\@arstrutbox}%
    \setlength\mylenb{#1\dp\@arstrutbox}%
    \rule[\mylenb]{0pt}{\mylena}}
\makeatother

\usepackage{bigstrut}
\newcommand{\urlwofont}[1]{ \urlstyle{same}\url{#1} }

\newcommand{\para}[1]{\noindent\textbf{#1}.}

\renewcommand{\vec}[1]{\mathbf{#1}}

\begin{document}

\begin{frontmatter}



\title{Tag Embedding Based Personalized Point Of Interest Recommendation System }

\author[addr1]{Suraj Agrawal\corref{correspondingauthor}}
\cortext[correspondingauthor]{Corresponding author.}
\ead{s.agrawal1993@gmail.com}
\author[addr2]{Dwaipayan Roy}
\ead{dwaipayan.roy@iiserkol.ac.in}
\author[addr1]{Mandar Mitra}
\ead{mandar@isical.ac.in}

  \address[addr1]{Indian Statistical Institute, Kolkata, India}
  \address[addr2]{Indian Institutes of Science Education and Research, Kolkata, India}

\begin{abstract}
E-tourism websites such as \emph{Foursquare}, \emph{Tripadvisor}, \emph{Yelp} etc. allow users to rate the preferences for the places they have visited.
Along with ratings, the services allow users to provide reviews on social media platforms.
As the use of hashtags has been popular in social media, the users may also provide hashtag-like \emph{tags} to express their opinion regarding some places.
In this article, we propose an embedding based venue recommendation framework that represents \emph{Point Of Interest} (POI) based on tag embedding and models the users (user profile) based on the POIs rated by them. 
We rank a set of candidate POIs to be recommended to the user based on the cosine similarity between respective user profile and the embedded representation of POIs. 
Experiments on TREC Contextual Suggestion data empirically confirm the effectiveness of the proposed model. 
We achieve significant improvement over PK-Boosting and CS-L2Rank, two state-of-the-art baseline methods. The proposed methods improve NDCG@5 by 12.8\%, P@5 by 4.4\%, and MRR by 7.8\% over CS-L2Rank.
The proposed methods also minimize the risk of privacy leakage. 
To verify the overall robustness of the models, we tune the model parameters by discrete optimization over different measures (such as AP, NDCG, MRR, recall, etc.). The experiments have shown that the proposed methods are overall superior than the baseline models. 
\end{abstract}


\end{frontmatter}


\section{Introduction}
\label{sec:intro}
Recent times have seen a growing interest in \emph{proactive} Information
Retrieval (IR) systems~\cite{DBLP:conf/ecir/BhatiaMA16} and related ideas
such as contextual suggestions~\cite{hashemi2016overview}, and zero-query
search systems. Such systems are designed to provide information that is
helpful to a user, given her current circumstances, without requiring the
user to submit an explicit query. Recommender Systems (RSs) that suggest
potentially interesting tourist attractions --- or \emph{Points Of
  Interest} (POIs) --- to travellers are examples of such proactive
systems. POI recommender systems~\cite{borras2014intelligent} seek to
address the following scenario. A tourist, while on a trip to a certain
place, looks for interesting things to do and sites to visit. The system
considers the traveller's current location and other factors (e.g., the
weather, whether the user is alone or with a group of friends). Based on its
prior knowledge about her preferences, it then recommends local POIs that
are most likely to be of interest to her.

\begin{figure}
  \centering
  \includegraphics[scale=0.3]{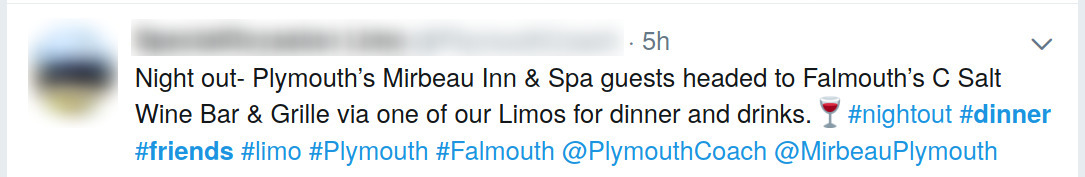}
  \caption{A sample tweet (anonymous) where the hashtags are indicative of the basic information contained in the tweet.}
  \label{fig:tweet1}
\end{figure}

The TREC Contextual Suggestion Track (TCST) that ran during
2012 -- 2016~\cite{DBLP:conf/trec/Dean-HallCKTV12,DBLP:conf/trec/Dean-HallCSKTV13,DBLP:conf/trec/Dean-HallCKTV14,DBLP:conf/trec/Dean-HallCKKV15,DBLP:conf/trec/HashemiKKCV16}
provided a comprehensive framework for investigating and evaluating RSs for
POIs. As with many other TREC tracks, the data and task definition for this
track evolved during its lifetime. In its final year (2016), an important
new feature was introduced: each POI was annotated with
user-assigned \emph{tags}\footnote{These were also called
  \emph{endorsements} at TREC.}. Tags are short descriptive labels assigned
to POIs by end-users. For example, the POI \emph{The Fitzwilliam Museum}\footnote{\url{https://www.fitzmuseum.cam.ac.uk/}} might be annotated
with the tags \ptag{museum} and \ptag{history}; similarly, \emph{Keen’s Steakhouse}\footnote{\url{http://www.keens.com/}}
could be tagged with \ptag{restaurant} and \ptag{pub}. More generally, the
use of tags has become a widespread practice in social media as a means of
quickly describing or classifying various items, like blog posts,
photographs, audio/video clips, etc. Hashtags in microblog posts are
possibly the most commonly encountered form of user-assigned tags.
Figure~\ref{fig:tweet1} shows an anonymous tweet describing an outing. The
tags assigned by the user to this tweet provide information about the
\emph{trip~type} (\ptag{nightout, dinner}), as well as the accompanying
\emph{type~of~group} ({\ptag{friends}}).
It is clear that user-assigned tags provide various types of information
(e.g., the \emph{category} of a POI --- \ptag{restaurant}, \ptag{museum})
that would be useful in the context of recommending POIs. Intuition
suggests that it should be possible to utilize this additional information
to provide better recommendations. Indeed, ``participants [at the TCST,
2016] showed considerable interest in using endorsements to improve
their contextual suggestion systems''~\cite{DBLP:conf/trec/HashemiKKCV16}.
In this study, we propose a method that leverages tags to improve the
quality of recommendations. Experimental results on the TCST 2016 data
suggest that the proposed method significantly advances the state of the
art.

Our proposed approach is primarily content-based. That is, if a
person 
has rated POIs $h_1, h_2, \ldots, h_m$ highly, but has given a low rating
to $l_1, l_2, \ensuremath{\ldots}, l_n$, then the predicted rating for POI
$p$ would depend on whether $p$ is more similar to $h_1, h_2, \ldots, h_m$
or to $l_1, l_2, \ensuremath{\ldots}, l_n$. From an IR perspective, one
natural way to quantify the similarity between two POIs would be to compute
the overlap between their textual descriptions. As indicated above, we
simply consider the set of tags assigned to a POI as its textual
description.
If we consider an operational
system (e.g., a mobile app) that actually implements our proposed method,
its interface could be designed to make it easy for an end-user to assign
tags to POIs. On the one hand, user-assigned tags are expected to be
high-quality content descriptors, somewhat analogous to keywords assigned
by subject experts. On the other hand, the number of tags assigned to a POI
will typically be small: for the TCST 2016 data, each POI was assigned 2.73
tags on an average. Thus, for this representation, the well-known
\emph{vocabulary mismatch problem}, which plagues any bag-of-words based
representation, is expected to be even more severe.

Word embeddings~\cite{mikolov2013distributed,pennington2014glove}
constitute a modern approach to tackling this problem. Broadly, an
embedding technique uses word-cooccurrence statistics from a (preferably large)
corpus of unlabelled textual data to map each word to a vector in a
low-dimensional, abstract space (the so-called \emph{embedding} space) in
such a way that the similarity between the vectors reflects the semantic
relationship between the corresponding words. Further, some simple vector
operations in the embedding space (e.g., addition and subtraction)
correspond nicely to linguistic notions such as conceptual combination, and
analogies. Embedding techniques have proven to be very useful in various
text processing tasks.

If embeddings are appropriately generated for POI tags, we expect that the
embeddings of `art galleries', `museums', and `poetry readings' will be
relatively close to each other, but relatively distant from the embeddings
of `parks', and `hiking trails'. Our approach, based on this intuition, can
be summarized as follows.
\begin{enumerate}
\item Embeddings are generated for all tags assigned to POIs.
\item For a particular POI, the embeddings of tags assigned to it are
  aggregated to obtain a representation of the POI in the embedding space.
\item Using Rocchio's algorithm~\cite{rocchio1971relevance} with the
  positive, negative and neutral ratings assigned by a traveller to various
  POIs, a profile is created for that traveller. This profile, also a
  vector in the embedding space, is expected to constitute an abstract, but
  high-level, description of the traveller's `type': whether the user prefers
  outdoor activities, or inclined towards the arts and
  literature, or whether of `food-drink-party' type, etc.
\item The profile of a traveller is matched against the vector
  representations of potential POIs, to be recommended, to assign scores. 
  Based on the scores, a ranked list of POIs are to be suggested to the user.
\end{enumerate}
Note that, in order to generate the embeddings in Step 1 above, we consider
the tags assigned by \emph{all} users to various POIs. 

In the next section, we summarize the contributions of this work.
Section~\ref{sec:related-work} discusses related work and
Section~\ref{sec:terminology} provides a quick review of the terminology
used in the TCST. The proposed method is described in detail in
Section~\ref{sec:approach}. Experimental results are presented and
discussed in Sections~\ref{sec:results} and \ref{sec:discussion}.
The article concluded in Section~\ref{sec:conclusion} listing an issue
that needs further investigation.

\section{Our Research Contribution} \label{sec:contrib}
\para{Research Objective}
Researches on venue recommendation surged its popularity among practitioners with the emerging usage of services like Foursquare, Tripadvisor, Yelp etc. across the globe.
The development of a productive recommender system would indeed be supportive to the users of services like these.
In this study, we propose to build a POI recommender system that would suggest places to visit considering the preferences of the user.

\para{Theoretical Contribution}
Given a fairly sized text collection containing well-formed sentences, word embedding techniques can produce a low-dimensional representation of words capturing the semantic relatedness of the terms.
The application of word embedding techniques for different text processing tasks have shown the effectiveness of the methods in terms of improving performances.
In this study, we argue that embeddings, when trained using a dataset consisting of informal tags without proper sentence structure, can also adept at capturing term relatedness. 

\para{Practical Implication}
The practical implication of the presented work lies in the empirical examination on whether training an embedding model using informal texts with casual labels could in effect capture the semantic regularities.
Experimentation on TREC Contextual Suggestion Track data
validates the significant predominance of the proposed embedding based approach over state-of-the-art techniques on the benchmark dataset.
Possible risk of most recommender systems is the concern of privacy breach that may happen due to the sharing of personal preferences with the server.
With this respect, the proposed approach is secure as the method can be designed to employ the user preferences for suggestions solely on the client's side without getting exposed to the server-side endpoint.
All codes and data that has been used in this article are available for reproducibility and further research purposes\footnote{\url{https://github.com/sagrawal1993/ContextualSuggestion}}.







\section{Related Work}\label{sec:related-work}

Researches on the development of recommender systems have been a popular choice among researchers in the information processing community.
The contextual information and preferences have been exploited for this purpose in the majority of the works~\cite{baltrunas2011matrix,adomavicius2011context,verbert2012context}. 
A number of collaborative filtering based approaches have been proposed for venue recommendation utilizing the behaviour of users with similar tastes~\cite{balabanovic1997fab,sarwar2001item,schafer2007collaborative,zheng2009wsrec,karatzoglou2010multiverse}.
The proximity and temporal information of the user are also used to recommend venues~\cite{hinze2003location, wijeratne2007local, roberts2009system, yuan2013time, catape}.
An extensive literature survey on general recommender systems is beyond the scope of this article. 
A comprehensive survey can be found in~\cite{bobadilla2013recommender} and~\cite{ricci2011introduction}. 
In the following section, we discuss some recommender systems which are developed for POI suggestion. Later in Section~\ref{sec:embeddings}, we focus on the related works in which embedding models are utilized for developing improved recommender systems for suggesting the place of attraction to users.


\subsection{Recommender Systems for e-Tourism}

Since the beginning of civilization, travelling has been the most popular recreational activity, and the recommendation of places has been a common practice among civilized people.
Early forms of the recommendations were the accumulated experience shared by the community. Modern recommender systems have evolved with the development of information technology. These systems have been used for different purposes, including recommending points of interests. 
The recommender systems developed for POI suggestion can be classified based on the adapted algorithms, user interfaces, and functionality provided by these systems~\cite{borras2014intelligent, RENJITH2020102078}. 
Lately, the increase in the use of mobile and smart devices leads to the development of various mobile recommender systems for tourists. The article by Gavalas et al.~\cite{gavalas2014mobile} classified mobile recommender systems in tourism along with highlighting the challenges in the field. The generated content using smart and mobile devices includes a lot of information in the form of rich media (images, videos, etc.), which requires efficient algorithms to be utilized for recommending places to travellers. A system based on media content has been proposed in~\cite{pliakos2014simultaneous}, which takes an image as input and produces a set of tourist places similar to the given image. 

With the advancement of GPS technology, location-based recommender systems have lately emerged for e-tourism. Park et al.~\cite{park2007location} propose a method, which uses the location of mobile devices along with time, weather, and user request to reflect user preferences and suggest appropriate places. Some recommender systems have been developed for specific proximity such as the \emph{myVisitPlanner}~\cite{refanidis2014myvisitplanner}, which is being used for residents of the region of northern Greece to plan their leisure, cultural activities during their stay in that area. Along with personal preferences, exploiting other tourist experiences could benefit while recommending new places. Yang et al.~\cite{yang2013itravel} have developed a system named \emph{iTravel}, which uses a peer-to-peer network to exploit the information and rating of other visitors of the same place. Recent development in technologies like augmented reality and 3D graphics have opened up a wide range of possibilities to add new functionality in tourist recommender systems. \emph{iGuide}~\cite{tsekeridou2014iguide} is one such system that provides interactive assistance to the users; it also offers a text to speech service along with augmented reality for the recommended place. A lot of personal data has been generated using smart devices which have been used by many venue recommender systems. The generated data may lead to privacy leak issues in these recommender systems. To minimize the risk of privacy leaks, Efraimidis et al.~\cite{efraimidis2016privacy} propose an approach by processing the information critical to users at the client end, utilizing the computation power of the user's device.

\subsection{Embedding based Recommender Systems} \label{sec:embeddings}

In recent years, deep learning has garnered considerable interest in computer vision and natural language processing, owing not only to the stellar performance but also to the attractive property of learning feature representations from scratch. The influence of deep learning is also prevalent, demonstrating its effectiveness when applied to information retrieval and recommender systems research. Zhang et al.~\cite{Zhang_2019} classified deep learning based recommender systems along with a summary of these methods. 
For recommendation tasks, \cite{poi2vec,geo-teaser} argued that places in the same geographical region should be assigned to similar representations as compared to the distant counterparts.
In contrast, the places can also be represented based on their types so that places with related genres are similar based on that representation.
Deep learning has also been applied for point of interest recommendation systems to capture different patterns. 
With the increase in social networking sites and location-based social networks, a lot of information has been generated regarding the point of interest in real-time. Hao et al.~\cite{hao2019real} propose a method to use such real-time information regarding POI along with intrinsic information. Convolutional Neural Network (CNN) has been used to capture intrinsic information along with multimodal embeddings to capture real-time information regarding POIs. As the user keeps visiting the places during a trip, their preferences change in real-time. It is challenging to capture such dynamic behaviour of the user. 
Utilizing the check-in sequence of the users and the characteristics of POIs by categories,~\cite{catape} proposes a POI embedding model CATAPE.
Graph-based POI embedding is proposed to capture such behaviour of the user~\cite{xie2016graph, xie2016learning}. 
The users may have short-term preferences regarding POIs. Manotumruksa et al.~\cite{manotumruksa2016modelling} propose a Contextual Recurrent Collaborative Filtering framework (CRCF) for short-term user preferences. In the sequence of visits to different POIs, it can be assumed that the latest visited POIs are more relevant to the user's preference than others. Tang et al.~\cite{tang2018personalized} propose a CNN based method to capture this pattern in the sequence of visited places by a user and recommend the future top N-POIs. It is also important to capture the transition behaviour from one POI to another by a user. SPENT~\cite{wang2019spent} captures such transition behaviours using word2Vec along with RNN. Most of the works consider the travel behaviour of the user, but the trip purpose is also an important feature that can be utilized. Chen et al.~\cite{chen2019trip2vec} propose a deep embedding based approach named Trip2Vec, to determine the trip purpose of the user. The preference of the users can be influenced by temporal factors such as day of the week (weekend/weekday) or time of the day (evening, night). Zhao et al.~\cite{zhao2017geo} propose a method that captures the changing behaviour of the user based on these temporal characteristics. 

There have been several studies that use the TCST dataset~\cite{hashemi2016overview} to develop contextual point of interest recommender systems.
These studies have covered a plethora of models and their variants while suggesting a point of interest based on user preferences and related context. Hashemi et al.~\cite{hashemi2016neural} propose a language model for user preference along with a word embedding based method to determine the boosting factors of endorsement terms. 
Some methods exploit word embedding to determine the venue and user's representation~\cite{manotumruksa2016modelling}. 
Content-based methods like weighted K-NN and rated Rocchio personalize query and their combinations have also been tried on this dataset~\cite{kalamatianos2016recommending}. 
User behaviours and opinions are being modeled using SVM classifier along with category and taste keyword information from other location-based social networking websites~\cite{aliannejadi2016venue}. 
Researchers have also tried to capture the reason behind the liking and disliking of a POI by the user and collaborative filtering based approaches are applied for the recommendations~\cite{Yang2015}. 
Most of these methods exploit travel history of the users. 
As another type of approach, Aliannejadi et al.~\cite{aliannejadiLinearCatRev} propose a model that is based on venue category and description keywords extracted from Foursquare tips, along with reviews from Yelp. 
Later, the work is extended by a method to determine the contextual appropriateness of a venue along with ways of incorporating different scores~\cite{aliannejadi2017venue}. They also propose a collaborative ranking framework to combine different similarity score~\cite{aliannejadi2018collaborative}. TCST dataset also contains user annotated tags, which are also being used by few of the methods. 
A probabilistic model is proposed in~\cite{aliannejadi2018personalized} that finds the mapping between user annotated tags and location's tag keywords. 
Their proposed model PK-Boosting~\cite{aliannejadi2018personalized} also use various learning to rank model to combine similarity scores. Considering the various types of methods developed for the POI suggestion tasks, Arampatzis et al.~\cite{arampatzis2018suggesting} compare and discuss the effectiveness, feasibility, efficiency, and privacy perspectives of content-based, collaborative filtering based and hybrid (combination of content-based and collaborative filtering based) methods. 

 The existing methods mostly use information like comments, categories, taste keywords, etc. associated with POIs for the recommendation of candidate POIs to a user. Other than this meta-information, tags are indicative of the category of a place based on its genre (\emph{bar, restaurant, museum,} etc.) as well as the type of visitors (friend, family, couple etc.) the place is mostly visited. 
 While considering tags, existing methods assume each tag to be independent; due to this, the semantic relatedness of the tags (like \ptag{spicy~food}, \ptag{local ~food} having similar intent) are lost. 
In this paper, we propose a set of POI recommendation methods that utilize the relatedness of the tags associated with POIs.
We employ \emph{word2vec}~\cite{mikolov2013distributed}, a word embedding model that captures the relationship between words.
While representing the users and ranking the candidate POIs, we use the embedding based approach in the Rocchio feedback framework~\cite{rocchio1971relevance}. 

\section{Terminology}
\label{sec:terminology}
We adopt the terminology used within the TREC Contextual Suggestion track.
For convenience, we summarize the main terms used in the rest of this
paper.

\smallskip
\noindent We use the terms \term{POI}, \term{Point Of Interest}, \term{venue,} and \term{place}
interchangeably to specify tourist attractions that are potentially of
interest to users.

\smallskip
\noindent \term{Tags} (or \term{endorsements}) refer to short descriptive
labels assigned to POIs by users. For example, a user may choose to assign
the tags \ptag{\{history, architecture\}} to \emph{The Colosseum}
(\url{https://www.coopculture.it/}). Depending on the implementation, users
may either have to choose and assign tags from a fixed taxonomy, or they
may be permitted to create their own tags.

\smallskip
\noindent A \term{profile} consists of a list of {POI}s rated by a
particular user, along with their ratings and {tags}. Some basic
information about the user (e.g., gender and age) is also included.
Table~\ref{tab:profile} shows an example of a user {profile}.
\begin{table}[h!]
  \centering
  \begin{tabularx}{0.95\linewidth}{llc}
    \texttt{Male, 29 years} \\\toprule
    \term{POI} & \term{Tags} & Rating \\\midrule
    \emph{Milan Cathedral - Duomo di Milano} & \ptag{\{history,~architecture\}} & 4 \\
    {\footnotesize\url{https://www.duomomilano.it/en/infopage/the-cathedral/53/}}
    \\[8pt]

    \emph{The Fitzwilliam Museum} & \ptag{\{history,~museum\}} & 3 \\
    {\footnotesize\url{https://www.fitzmuseum.cam.ac.uk/}} \\[8pt]
    
    \emph{The Temple Bar} & \ptag{\{pub,~beer,~bar-hopping\}} & 1 \\
    {\footnotesize\url{https://thetemplebarpub.com/}}
    \\[8pt]

    \emph{Keen's Steakhouse} & 
    \ptag{\{pub,~restaurants\}} & 2 \\
    {\footnotesize\url{http://www.keens.com/}} \\\bottomrule
  \end{tabularx}
  \caption{An example of a user \term{profile}}
  \label{tab:profile}
\end{table}

\smallskip
\noindent A \term{context} describes the circumstances in which
recommendations are to be generated. For this study, a context consists of
\begin{itemize}
\item a \emph{destination city} (e.g., Amsterdam), which gives the user's
  location;
\item a \emph{trip~type} (e.g., leisure, work);
\item a \emph{trip~duration} (e.g., weekend, one week, two weeks);
\item \emph{type~of~group} the person is traveling with (e.g., family,
  friends, colleagues); and 
\item \emph{season} of the trip (Spring, Summer, etc.).
\end{itemize}

\smallskip
\noindent A \term{request} consists of a $\langle$ context, profile
$\rangle$ pair, and corresponds to a specific request for recommendations
for the given profile in the given context. 

POIs and requests may thus be regarded as the TCST's analogue of documents
and queries respectively, in the sense that a Contextual Suggestion system
is supposed to respond to a request with a ranked list of POIs. 



\section{Tag Based POI Recommendation} \label{sec:approach}
As explained using various examples in the preceding sections, tags
encapsulate high-level attributes of a POI such as its \emph{category}, and
the \emph{trip type} for which it would be appropriate. Thus, tags assigned
to POIs are analogous to the keywords assigned as content descriptors to a
piece of text, and the semantic overlap between the tag sets assigned to two
POIs may be considered to be a measure of their similarity.
In this section, we will be using examples of tags from the TREC CS track 2016 data~\cite{hashemi2016overview}, which we have used for evaluating our proposed methods, and have described in detail in Section~\ref{sec:results}.

In traditional IR models, keywords (or tags) correspond to orthogonal dimensions, or equivalently, to sparse --- or
\emph{one-hot} --- vectors in a high-dimensional space (or something similar
in various formalisms). According to such a representation, the
tags \ptag{parks} and \ptag{outdoor activity} are as unrelated as the tags
\ptag{outdoor activity} and \ptag{art galleries}. Clearly, we need to
represent tags in a way that better captures the semantic relation between
tags. 
Word embeddings try to achieve this objective by projecting each word of a
given vocabulary to a \emph{dense} vector in a low-dimensional abstract space
in such a way that strongly related words generally correspond to `similar'
vectors.


\subsection{Embedded representation of tags}\label{subsec:wordembed}

Embedding strategies operationalize the well-known linguistic principle
that a word is known \emph{by the company it keeps}, i.e., words that are
strongly semantically related occur very often in similar \emph{contexts},
where a {context} is typically defined as a window of contiguous words
within a sentence. We hypothesize that a similar principle applies to tags
as well. 
For example, suppose the POI `Universal Studios
Singapore'\footnote{\url{https://www.rwsentosa.com/en/attractions/universal-studios-singapore}}
is tagged with \ptag{family-friendly}, \ptag{outdoor-activities}, and
\ptag{parks}, and another POI, `Sentosa
Boardwalk'\footnote{\url{https://www.sentosa.com.sg/en/}} has tags
\ptag{citywalks}, \ptag{family-friendly}, and \ptag{outdoor-activities}.
This suggests that the tag \ptag{parks} and \ptag{citywalks} may be
related.
Accordingly, we create a `sentence' out of all the tags assigned to a particular POI by users.
Different orderings (or permutations) of each POI's
tag-set may yield additional sentences. 
Tag embeddings can then be created in the usual way from this corpus of sentences.

Once the tags have been mapped to an embedding space, we use this
representation to also map POIs and user profiles into the embedding space.
The following sections provide details of how this is done. Finally,
Section~\ref{sec:ranking} describes how embeddings can be used to rank
candidate POIs for a particular user in decreasing order of predicted
attractiveness.

\subsection{Modeling POIs}
\label{model_pois}
A
representation of the POI in the embedding space may be obtained by
aggregating the embeddings computed for the tags assigned to that POI.
Since the number of tags per POI is generally small, we use a simple
summation. Various other aggregation strategies have been proposed in the
literature~\cite{zhang-ecir18,boom-2016}. Two obvious alternatives to computing the sum are
computing the centroid and using the Doc2Vec model~\cite{lau2016empirical}.
Our preliminary experiments suggested that a simple summation works as well
as, or better than, these alternatives. For the results reported in this
article, therefore, we compute the embedding for POI $P$ as
\begin{equation}
  \vec{P}=\sum_{tg \in \mathit{TG}(P)}\vec{tg}  \label{eq:poi_vec}
\end{equation}
where \(\mathit{TG}(P)= \langle tg_1, tg_2, \ldots, tg_k \rangle\) is the
tags assigned to $P$.

\subsection{Modeling users}
\label{model_users}
A person may want to visit similar places in the future based on their past experience or may avoid certain places entirely. 
When the person is intrigued by the location, these two most distant preferences (either positive or negative) are expressed. 
Generally, when the user is unprejudiced by the place, a neutral preference (or feedback) is expressed. 
The users are modeled based on the feedback for the visited places.
The modeling is carried out based on three dimensions of preferences of the user: regardless of whether the user enjoyed, loathed, or has no strong assessment on the place; the representations are respectively called positive profile, negative profile, and neutral profile.

The preferences of the users are rated in a scale of 0 (highly disliked) to 4 (profoundly enjoyed); we consider the neutral preference with rating value of 2.
Further, the places with a rating greater than 2 are considered as relevant/positive or liked, while a rating of less than 2 are treated as non-relevant/negative or disliked by the user. 
Based on the way the venue ratings are considered, user profiles can be modeled in two ways.
In the \emph{unweighted} variant, all the relevant ratings (3, 4) are assigned equal positive importance; likewise equal negative importance are assigned for non-relevant ratings (0, 1). 
Alternatively, in the \emph{weighted} variant, higher importance are assigned to strongly relevant profiles (i.e., those rated 4) than to weakly relevant profiles (i.e., those rated 3); likewise, for strongly and weakly non-relevant (negative) profiles.
The description of the user modeling variants are formally discussed as follows:

\begin{itemize}
\item{\textbf{Unweighted:} }
In this representation of users, equal significance are given to POIs that are rated either as strongly positive (rating 4) or as weakly positive (rating 3);
finally, the \emph{positive profile} is made with those POIs that are rated by the user with a positive rating. 
A positive profile for the user is formulated by composition of the vectors (addition) corresponding to POIs relevant to the user (i.e. rated either as 3 or 4). 
Formally for user $u$, let the relevant POIs be: \( prof^+(u)=<P_1,P_2,\ldots> \). 
Then, the positive profile \emph{u}, denoted by $\overrightarrow{prof_{uw}^+(u)}$ is presented in Equation~\ref{eq:prof_unweighted}.
\begin{equation}
\overrightarrow{prof_{uw}^+(u)}=\frac{\sum_{P\epsilon prof^+(u)}\vec{P}}{|prof^+(u)|} \label{eq:prof_unweighted}
\end{equation}

In the same manner, we can create a \emph{negative profile vector} \(\overrightarrow{prof_{uw}^-(u)}\), and a \emph{neutral profile vector} \(\overrightarrow{prof_{uw}^o(u)}\) which are formally presented in Equations~\ref{eq:prof_unweighted_neg} and \ref{eq:prof_unweighted_neu} respectively.

\begin{equation}
\overrightarrow{prof_{uw}^-(u)}=\frac{\sum_{P\epsilon prof^-(u)}\vec{P}}{|prof^-(u)|} \label{eq:prof_unweighted_neg}
\end{equation}

\begin{equation}
\overrightarrow{prof_{uw}^o(u)}=\frac{\sum_{P\epsilon prof^o(u)}\vec{P}}{|prof^o(u)|} \label{eq:prof_unweighted_neu}
\end{equation}

\begin{table}[h]
\begin{center}
    \begin{tabular}{| l || c | c | c | c | c |}
   \hline
        Ratings given by user &  0 &  1 & 2 & 3 & 4 \\
         \hline
        Scaled rating &  -3 & -2 & 1 & 2 & 3 \\
    \hline
    \end{tabular}
\end{center}
\caption{The scaled ratings for the POIs.}
\label{table:rating}
\end{table}

\item{\textbf{Weighted:} }
Strongly positive POIs (rating 4) are more preferred by the user than weakly positive POIs (rating 3). 
In the unweighted representation, we are not utilizing this strong opinions.
An obvious way of incorporating this is by considering extra importance to strongly positive (or negative) preferences than weakly positive (or negative) ones, and model the positive (and negative) profiles.
The fondness for a place to the user is determined by the rating given by her. 
In order to have the weights scaled with the corresponding likes and dislikes associated respectively with the positive and negative ratings, we change the ratings following Table~\ref{table:rating}. Note that, the scaled ratings would help to fairly distribute the weights for the different POIs accordingly to the preferences; same weights but different directions are assigned to weakly positive and weakly negative (3 and -3), as well as strongly positive and strongly negative ratings (4 and -4).
Given the relevant POIs to user $u$ are \( prof^+(u)=<P_1,P_2,\ldots> \),
a \emph{positive profile vector} can be formulated following Equation~\ref{eq:prof_weighted}. 
\begin{equation}
\overrightarrow{prof_w^+(u)}=\frac{\sum_{P\epsilon prof^+(u)}\vec{P}*rating(P)}{|prof^+(u)|} \label{eq:prof_weighted}
\end{equation}
Similarly, we can form \emph{negative profile vector} \(\overrightarrow{prof_w^-(u)}\) and neutral profile vector \(\overrightarrow{prof_w^o(u)}\) by considering the negative and neutral ratings given to POIs by the user. 
Note that, for a POI with neutral rating (scaled value $1$ in Table~\ref{table:rating}), the scaling would result in the same weight for the weighted profile as that of the unweighted one; thus essentially a neutral rating would not be adding any extra value to either of weighted or unweighted profiles.

\end{itemize}

It is apparent that a user would rate some places as positive, negative, and some as neutral following the preferences.
Based on these preferences, a positive, a negative and a neutral profile vectors can be constructed for each user in the way discussed earlier in this section.
Considering these different signals, we formalize a user specific \emph{user profile vector} using the idea of Rocchio model ~\cite{rocchio1971relevance}.
In the document retrieval scenario, Rocchio feedback method works in Vector Space Model (VSM) where the initial query vector $\vec{Q}_i$ is modified to $\vec{Q}_m$ based on the centroid of the set of relevant ($D_R$) and non-relevant ($D_{NR}$) documents. The model is presented in Equation~\ref{eq:rocchio}.
\begin{equation}
    \vec{Q}_{m} = \alpha \frac{1}{|D_R|}\sum_{\vec{d_i}\in D_R}{\vec{d_i}}
    + \beta  \vec{Q}_{O}
    - \gamma \frac{1}{|D_{NR}|}\sum_{\vec{d_j}\in D_{NR}}{\vec{d_j}} \label{eq:rocchio}
\end{equation}
The parameters $\alpha$ and $\gamma$ in Equation~\ref{eq:rocchio} are associated with the weights for set of relevant and non-relevant documents (respectively $D_R$ and $D_{NR}$)
while $\beta$ corresponds to the weight for the original question. 
%
We create user vectors following the same methodology from the corresponding positive (relevant), negative (non-relevant) and neutral (original) profile representations. 
Formally, following the exposition of Rocchio model, a linear combination of positive, negative and neutral profile vectors are taken to create an overall user profile vector.
Mathematically, the \emph{unweighted user profile} (corresponding to Equation~\ref{eq:prof_unweighted}~-~\ref{eq:prof_unweighted_neu}) is defined as Equation~\ref{eq:unweighted-poi-rocchio}. 
\begin{equation}
    \overrightarrow{prof_{uw}(u)}=\alpha*\overrightarrow{prof_{uw}^+(u)}
    + \beta*\overrightarrow{prof_{uw}^o(u)}
    - \gamma*\overrightarrow{prof_{uw}^-(u)} \label{eq:unweighted-poi-rocchio}
\end{equation}
For the weighted user profile, the final vector is formulated following Equation~\ref{eq:weighted-poi-rocchio}.
In contrast to the Rocchio model (Equation~\ref{eq:rocchio}), $\alpha$, $\beta$ and $\gamma$ are parameters of the profile vectors indicating respectively the weights associated with positive, neutral and negative profiles in Equations~\ref{eq:unweighted-poi-rocchio} and~\ref{eq:weighted-poi-rocchio}.

\begin{equation}
    \overrightarrow{prof_w(u)}=\alpha*\overrightarrow{prof_w^+(u)}
    + \beta*\overrightarrow{prof_w^o(u)}
    - \gamma*\overrightarrow{prof_w^-(u)} \label{eq:weighted-poi-rocchio}
\end{equation}

\subsection{Ranking POIs} 
\label{sec:ranking} 
Once all the POIs and user profiles are represented as vectors in the embedding
space, the candidate POIs for a particular user profile can be easily ranked
based on some measure of vector similarity. In this work, we use the
\emph{cosine similarity} between a POI vector \(\vec{P}\)
(Equation~\ref{eq:poi_vec}) and a user profile vector, either the
unweighted version, \(\overrightarrow{\mathit{prof}_{uw}(u)}\),
(Equation~\ref{eq:unweighted-poi-rocchio}), or its weighted variant
\(\overrightarrow{\mathit{prof}_{w}(u)}\)
(Equation~\ref{eq:weighted-poi-rocchio}) to rank the POIs based on their similarities with the profile vectors.


\section{Experimental Setup} 
\label{sec:results}

\subsection{Dataset} \label{subsec:dataset}
We use the dataset provided by the \emph{TREC Contextual Suggestion (CS) track}\footnote{\url{https://trec.nist.gov/data/context.html}} for empirical evaluation of our proposed model.
The overall objective of the track was the development and investigation of techniques for addressing ``complex information needs that are highly dependent on context and user interests''~\cite{hashemi2016overview,DBLP:conf/trec/Dean-HallCKTV12,DBLP:conf/trec/Dean-HallCSKTV13,DBLP:conf/trec/Dean-HallCKTV14,DBLP:conf/trec/Dean-HallCKKV15}. During its tenure, the track focused mainly on the problem of recommending tourist attractions, or \emph{Points of Interest} (POIs), to travelers. 

Table~\ref{table:dataset} presents some basic statistics about the datasets
used in the TCST during 2015 and 2016. Recall from the Introduction that
the primary objective behind this study was to explore how tags may be used
to improve recommendation quality. In 2015, tags were introduced for the
first time. For any POI rated by a user in their profile, the user was
permitted (but not required) to annotate the POI with tags to indicate why
the user liked the particular POI. Row 9 of Table~\ref{table:dataset} shows
that a total of 11,400 ratings (across all profiles) were obtained in 2015.
Of these, only 6,599 (less than half) were assigned tags.
More importantly, the set of candidate POIs comprising the target `document
collection' were not assigned any tags at all. Thus, the 2015 dataset could
not be used for evaluating our method.
In 2016, tag information was included in a more systematic
way.\footnote{\url{https://sites.google.com/site/treccontext/trec-2016}} A
large proportion (4791/5599, see Row 10 of Table~\ref{table:dataset}) of
the candidate POIs in the target `document collection' were assigned tag(s)
(this was in addition to the 2273 POIs that were rated and tagged by users
during profile construction --- cf.\ Row 9, Table~\ref{table:dataset}).
Thus, the 2016 task permitted groups to use tag-based retrieval or ranking
strategies for the first time. We use the dataset from 2016 to evaluate our
methods, and to compare its performance with various baselines.
A comprehensive discussion on the TREC 2016 CS track dataset can be found in~\cite{hashemi2016overview}.

  \begin{table}[]
    \centering
    \begin{tabular}{clp{3cm}r}
      \hline
      Row & & TREC CS 2016 & TREC CS 2015 \\
      \hline
      1 &  Number of requests ($\approx$ number of queries)     &   442  &    221   \\
      2 &  Number of requests evaluated by TREC          &   58      &   211     \\
      3 &  Number of POIs ($\approx$ size of doc.\ collection)      &  18,808    &   8,794   \\
      4 &  Number of users       &  238    & 209       \\
      5 &  Number of distinct users in set of evaluated requests  &    27   &   209     \\
      6 &  Number of POIs rated per user  & 30 or 60 & 30 or 60 \\
      7 &  Number of distinct POIs rated across all users  & 60\footnote{All users rated the same set of 60 POIs.} & 4102 \\
      8 &  Number of  candidate POIs returned (or reranked)  per request & 79-119 (AVG 96.5) & 30\\
      9 &  Number of POIs with tags across all profiles & 2273/2310 & 6599/11400\\
      10 & Number of candidate POIs with tags in evaluated requests & 4791/5599 & 0/6330 \\
      11 & Number of unique tags & 150 & 186 \\
      \hline
    \end{tabular}
    \caption{TREC Contextual Suggestion track dataset 2015, 2016}
    \label{table:dataset}
  \end{table}

\begin{figure}[h!]
\centering
\fbox{\includegraphics[scale=0.35]{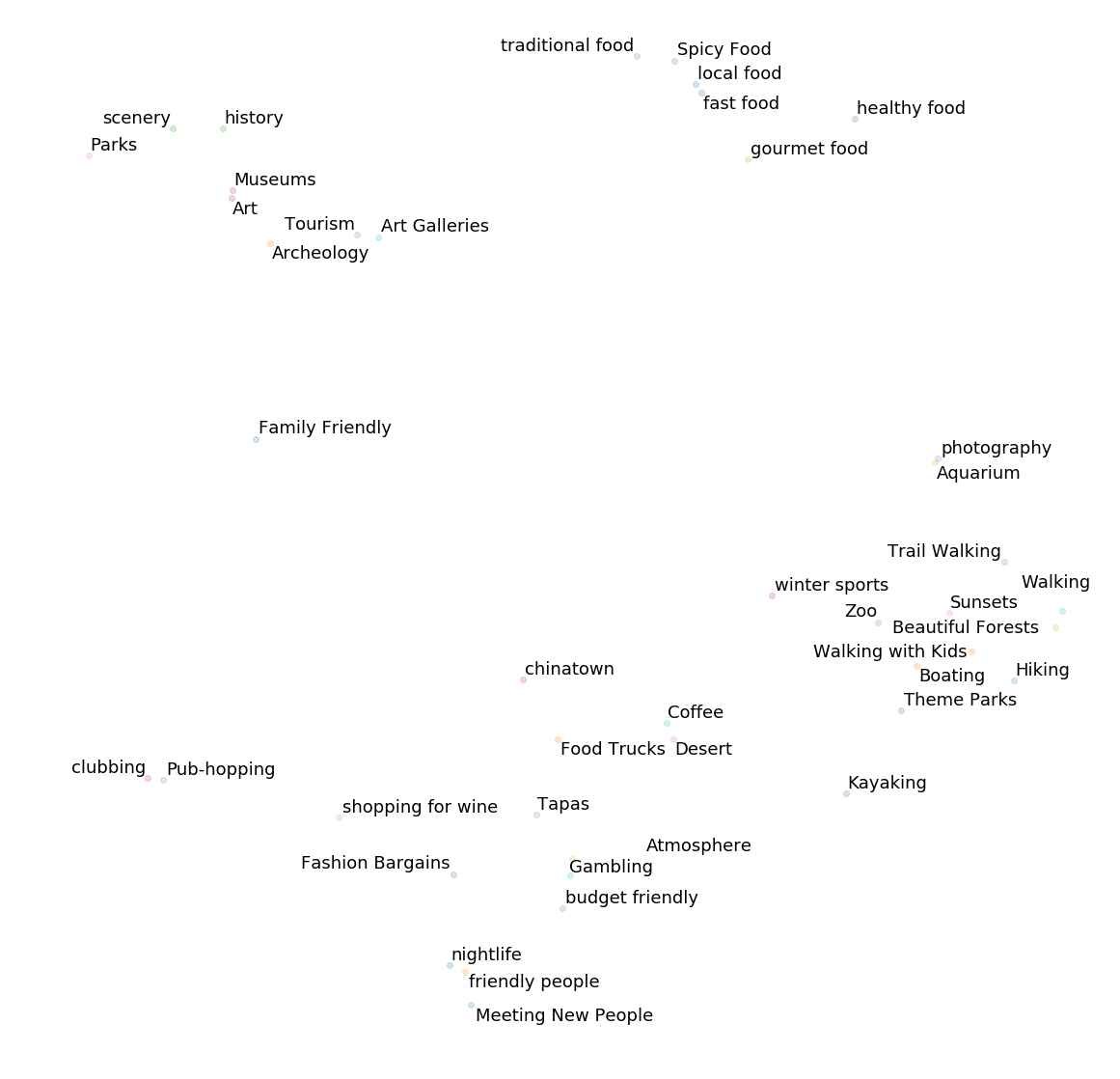}}
\caption{A two dimensional projection of the embedded vector representations corresponding to the tags using \emph{\href{https://lvdmaaten.github.io/tsne/}{t-SNE}}~\cite{tsne}. }
\label{fig:method}
\end{figure}

\subsection{Training the embeddings} 
\label{subsubsec:train-embed}
It has been a common practice among researchers to use pre-trained embedding models for retrieval tasks~\cite{diaz-etal-2016-query,zamani2018sigir,zheng2015}. 
However, previous studies have reported that training the embedding model using an in-domain collection could result
in better performance as compared to using pre-trained models~\cite{roy2018cikm}. 
Accordingly, no pre-trained models were used in our experiments; instead, we used data from the TREC CS track itself to train the embedding model as described below.

The quality of the trained embedding model also depends on the quantity of data used for training.
To observe the variation in performance, we make two embeddings 
$(i)$ using only the TREC CS track 2016 dataset, and $(ii)$ combining both TREC 2015 and 2016 CS track datasets.
For representing the tag terms in an embedded space, any standard embedding models (\cite{mikolov2013distributed,pennington2014glove,bert}) can be applied. 
In this study, we have used \emph{word2vec}~\cite{mikolov2013distributed} for learning the vector representations of tag terms, however we believe that similar performance would be observed in case similar embedding techniques are used~\cite{zuccon2015adcs,roy2018cikm}.
To train the word2vec model, we consider all the tags assigned to a particular POI by the users and form a sentence. 
As an example, consider a POI has been assigned the tags: \ptag{Tourism}, \ptag{Culture}, \ptag{Family~Friendly}, \ptag{Food}, \ptag{Entertainment}. 
Then the constructed sentence associated with this place will be `\emph{tourism culture family-friendly food entertainment}'. 
Thus we create a set of sentences for each of the POIs in the dataset.
This list of sentences is used as the text dump to train the word2vec model.
%

The word2vec model is trained using the continuous bag-of-word (CBOW) model with the window size (parameter \texttt{window} as specified in word2vec) varied in the range of 3 to 10 in steps of 1. 
The minimum term count (\texttt{min-count}) is varied from 2 to 7 in steps of 1. 
As reported in~\cite{zuccon2015adcs, roy2018cikm}, we observed similar performance with insignificant difference after varying these parameters.
In this article, we report the results obtained using \texttt{window} size and \texttt{min-count} respectively set to 5 and 3.
As there are fewer number of unique tag terms (specifically $150$) in the dataset, the embedding dimension (\texttt{size}) is set to $9$ after varying it from $5$ to $20$ in unit steps.
Note that, the embedding dimension used here is considerably smaller than the normal text retrieval settings where the dimension ranges between $100$ to $500$~\cite{ganguly2015,grbovic2015sigir,roy2019ipm,roy2016cikm,mitra2017}.
As the training data is restricted in terms of size, we expect to get a better representation when more iterations are applied while training (similar to~\cite{diaz-etal-2016-query}). 
The number of iterations (parameter \texttt{iter} as specified in word2vec) are therefore varied in the range $\{5, 10, 50, 100, 200, 500, 1000, 1500, 2000\}$ and finally set it to $1000$ based on the initial performance.
All other parameters associated with word2vec are set to their default values.
The variation in performance, when the number of iterations and the dataset sizes are varied, is discussed in Section~\ref{sec:discussion}.


The two-dimensional projection of the trained model using t-SNE~\cite{tsne} in presented in Figure~\ref{fig:method}.
In accordance with our intuition, it can be seen from the figure that tags with similar semantics are located in close proximity to each other.
For an example, notice that related tags like \ptag{art~galleries, history, tourism} and \ptag{art} are close to \ptag{museums} in the abstract space. Similarly, tags related to foods are close to each other in the two-dimensional projection of the embeddings. 
This empirically verifies the ability of embeddings, specifically the word2vec model~\cite{mikolov2013distributed} in our case, in capturing the relationship between tags. 

\subsection{Sparse one-hot encoded representation of tags} \label{subsec:onehot}
As presented in the Section~\ref{sec:approach}, the proposed approach utilizes word embedding that creates a dense representation for each of the POIs which enables us to capture tag similarity and to model the preferences of a user based on their likes and dislikes.
The positive and negative preferences (respectively likes and dislikes) are then applied in a Rocchio feedback framework to make an user model.
While creating the dense representations, the embedding model considers the relationships between the tags.
To verify the functionality of the dense representation, we conduct another set of experiments where the dense tag representation is replaced 
with a sparse one-hot encoded vector.
In this representation, the number of unique tags in the dataset would be the size of the vector, which is 150 (as mentioned in Table~\ref{table:dataset}) where each dimension in the vector represents an individual tag information.
In the representation of a tag using a one-hot vector, the dimension corresponding to that tag will contain 1 and the remaining dimensions will have the value 0.
Note that, in contrast with training of the word embedding model, there are no parameters associated with making this one-hot encoded representation of the tags.
Also the tags are considered orthogonal to each other, without having any relationships.

\subsection{Tuning parameters} \label{sec:parameters}

The proposed tag-embedding based recommendation model has 3 parameters, $\alpha$, $\beta$ and $\gamma$ of Equations~\ref{eq:unweighted-poi-rocchio} and \ref{eq:weighted-poi-rocchio}, that weigh the positive, neutral and negative profile representations respectively.
Rather than setting these parameters heuristically, we adapt a discrete optimization method considering different metric values as an optimization function to locate the optimum settings for these parameters.
To avoid overfitting the performance based on a single metric, we select various measures such as $NDCG@5$, $P@5$, and $MRR$ of the recommendation system for optimization and to analyze the result.

\para{Optimization function score}
Consider the \term{profile} of a user $u$ which contains the rated POIs as well as tags/endorsements given to those places (see Table~\ref{tab:profile}). 
Depending on the past experience, if $u$ has rated \emph{Milan Cathedral - Duomo di Milano}\footnote{\url{https://www.duomomilano.it/en/infopage/the-cathedral/53/}} (associated tags \ptag{\{History, Architecture\}}) with $4$, \emph{{The Fitzwilliam Museum}}\footnote{\url{https://www.fitzmuseum.cam.ac.uk/}} (associated tags \ptag{\{History, Museum\}}) with $3$ and \emph{The Temple Bar}\footnote{\url{https://thetemplebarpub.com/}} (associated tags \ptag{\{Pub, Beer, Bar-hopping\}}) with $1$, the corresponding ratings will connote the preferences of that user. 
The ratings provided by the user, which reflect the preferences, is utilized to model a user as discussed in Section~\ref{model_users}.
The unweighted and weighted user models (respectively Equations~\ref{eq:unweighted-poi-rocchio} and~\ref{eq:weighted-poi-rocchio}) contain three parameters $\alpha$, $\beta$ and $\gamma$ that control the positive, neutral and negative profile representations respectively.
Following the discussion in Section~\ref{model_pois}, a vector for each of the POIs in the user \term{profile} (e.g. \emph{Milan Cathedral - Duomo di Milano}, \emph{The Fitzwilliam Museum}, \emph{The Temple Bar}, \emph{Keen’s Steakhouse}) can also be generated.
Finally the ranking of these POIs can be generated by using cosine similarity between user profile vector and the POI vectors (as discussed in Section~\ref{sec:ranking}).
This ranking can be evaluated based on the ratings as given in the user \term{profile}.
Here, the ratings for the POIs (last column in Table~\ref{tab:profile}), available as part of \term{profile}, are considered as the ground truth.
Note that, we are \emph{not} using the relevance judgement of the request queries, containing the candidate POIs to be ranked, for tuning the parameters.
We evaluate the prediction of the \term{profile} POIs based on the evaluation metrics, specifically NDCG@5, P@5 and MRR; these performance measures can also be used as the optimization function scores.


The parameter associated with the neutral preference of a user ($\beta$ in Equations~\ref{eq:unweighted-poi-rocchio}, and \ref{eq:weighted-poi-rocchio}) is set to $1.0$ to keep it uniform for all users. 
While tuning the other two parameters (namely $\alpha$, and $\beta$) on the \term{profile} data, we consolidate two strategies to find the optimum values.
In the first set of experimentation, we apply \emph{genetic algorithms}~\cite{deb2002fast} to find the optimal range of the parameter values which is observed to be in the range $(-8.0, 8.0)$.
We then exhaustively explore further using \emph{grid search} in the subspace selected by the genetic algorithm varying the values in an interval of $0.2$.
The corresponding values of $\alpha$ and $\gamma$ are selected that improves the performance prediction in terms of NDCG, P@5 and, MRR when evaluated based on the user \term{profile}. 
After optimizing the metrics to get the optimal parameter settings, we explore two ways of assigning the optimal values of $\alpha, \beta$, and $\gamma$ for each \term{profiles} while performing the prediction of the preferences: 

\subsubsection{Same parameter for all user model}\label{subsubsec:sameparam}
There are multiple users, each representing their own preferences. 
In the first approach of assigning the parameter values, we set the same values for all the user models.
To find this value of the parameters ($\alpha, \beta$, and $\gamma$), we apply genetic algorithms and grid search (as discussed earlier) on the user \term{profiles}, and then maximize the average optimization score over all users. 

\subsubsection{Unique parameters per user model}\label{subsubsec:uniqparam}
If the same parameter values ($\alpha, \beta$, and $\gamma$) are applied while creating the model for all the users, the models may have become generalized. 
Another way of making the user models would be to use separate weights for each of the users depending on their positive, neutral, and negative preferences. 
Based on individual preferences, this can be done by tuning the parameters ($\alpha, \beta$, and $\gamma$) uniquely per user and give weights accordingly. 
Further, to find the value of parameters, we apply grid search as mentioned above to maximize the optimization score for individual users.

\subsection{Evaluation measures}

In this study, we evaluate the proposed recommendation methods and compare with the baseline models using three evaluation metrics, specifically NDCG@5, P@5, and MRR. 
Note that, these metrics are also the official metrics used in the TREC Contextual Suggestion track~\cite{DBLP:conf/trec/ManotumruksaMO16}. 
%

\subsection{Experimentation} \label{sec:experiments}

Following the discussion in Section~\ref{model_users}, a user model can be generated in an abstract vector space in two ways: an unweighted version where strongly positive ratings are considered with equal importance with weakly positive ratings (similarly strongly and weakly negatives are considered with equal importance) as in Equation~\ref{eq:unweighted-poi-rocchio}, and the weighted version that considers the actual ratings for the places provided by the users while formulating the preference vector (Equation~\ref{eq:weighted-poi-rocchio}).
Also the three parameters of the user model ($\alpha$, $\beta$ and $\gamma$ of Equations~\ref{eq:unweighted-poi-rocchio} and~\ref{eq:weighted-poi-rocchio}) can either be tuned in a per user basis, or the same settings can be applied for all users.
Based on this two ways of representing the user profiles and setting parameters, we experiment with the following variants:
%
\begin{enumerate}
    \item \textbf{WUPSame} (Weighted User Profile with Same Parameter for all users):
We create user model using the weighted representation (Equation~\ref{eq:weighted-poi-rocchio}), while same parameter values ($\alpha$, $\beta$, and $\gamma$) for all profiles are considered. 

\item \textbf{UnWUPSame} (Unweighted User Profile with Same Parameter for all users):
In this variant, the unweighted representation are used for 
the user (Equation~\ref{eq:unweighted-poi-rocchio}); further same parameter values ($\alpha$, $\beta$, and $\gamma$) for all user models are considered. 

\item \textbf{WUPUniq} (Weighted User Profile with Unique Parameter for each user):
A per user parameter tuning is performed in this variant while the weighted representations of user model (Equation~\ref{eq:weighted-poi-rocchio}) is used for modeling the user.

\item \textbf{UnWUPUniq} (Unweighted User Profile with Unique Parameter for each user):
As the fourth variant, we use the unweighted representation for the user model (Equation~\ref{eq:unweighted-poi-rocchio}) and the parameters are tuned separately for each user.
\end{enumerate}

Similar alternatives to model POIs and users can also be adapted with the sparse, one-hot encoded representation of the tags (as discussed in Section~\ref{subsec:onehot}).
We devise additional methods that utilize the one-hot vectors of the tags, replacing the dense representation.
Specifically, we formulate four models 
\textbf{WUPSame$_{01}$}, \textbf{UnWUPSame$_{01}$},
\textbf{WUPUniq$_{01}$} and \textbf{UnWUPUniq$_{01}$}
which are similar to WUPSame, UnWUPSame,
WUPUniq and UnWUPUniq respectively, with the one-hot
representations are being used instead of the word2vec embedding.

\subsection{Baseline models} \label{sec:baseline}
We compare the proposed methods with a number of state-of-the-art baselines.
We choose DUTH~\cite{kalamatianos2016recommending} that uses a K-NN classifier to select the ratings of candidate POIs and generate a query following the Rocchio algorithm.
UAmsterdamDL~\cite{hashemi2016neural} learns a language model for each user profile and POI utilizing the tags, their embedded representations (presented in the POI description); final ranking of the POIs is performed based on the KL divergence score between the POIs and profile. 
USI~\cite{aliannejadi2016venue} generates multiple scores based on reviews of the POIs rated by the user, normalized frequency score of the category and taste keywords, and score of the context appropriateness; it combines these scores using linear interpolation before ranking the POIs.
Venue appropriateness prediction (CS-L2Rank)~\cite{aliannejadi2017venue} is an extension of USI, that uses learning to rank model to combine the scores, which further improves the performance.
The work is further extended in PK-Boosting~\cite{aliannejadi2018personalized} which finds the mapping from taste keywords to tags (assigned by the assessor) corresponding to a POI using the maximum likelihood and sequence labelling techniques.

We have selected the baselines as the working principle of these methods are, in some aspects, similar to the proposed methods. 
DUTH~\cite{kalamatianos2016recommending} uses the modified Rocchio algorithm to generate a query representing the user and search for candidate POIs in the user's context. 
Similar to DUTH~\cite{kalamatianos2016recommending}, we utilize the Rocchio relevance feedback algorithm for user representation.
PK-Boosting~\cite{aliannejadi2018personalized} uses tags to refine the taste keywords. Word embedding is also being used by UAmsterdam~\cite{hashemi2016neural}, but for finding the term representation, which is used to determine the relevance of terms to a tag. %
Our approach only uses tag information associated with POIs, while PK-Boosting used it as additional information to refine the taste keywords corresponding to a POI (gathered from location-based social networking sites). Also, PK-Boosting considers tags to be independent while our method utilizes word embedding to capture the relationship between tags.

\begin{table}
\begin{center}
    {\begin{tabular}{ l l l l l l l }
  \hline
     & NDCG@5 & $\Delta(\%)$ & P@5 & $\Delta(\%)$ & MRR & $\Delta(\%)$ \\
   \hline
        UAmsterdamDL~\cite{hashemi2016neural} & 0.2824 & - & 0.4448 & - & 0.5924 & -  \\
        LinearCatRev~\cite{aliannejadiLinearCatRev} & 0.3213 & - & 0.4897 & - & 0.6284 & - \\
        USI5~\cite{aliannejadi2016venue} & 0.3265 & - & 0.5069 & - & 0.6796 & -  \\
        DUTH\_rocchio~\cite{kalamatianos2016recommending} & 0.3306 & - & 0.4724 & - & 0.6801 & - \\
        PK-Boosting~\cite{aliannejadi2018personalized} & 0.3526 & - & 0.5310 & - & 0.6800  & - \\
        CS-L2Rank~\cite{aliannejadi2017venue} & 0.3603 & - & \textbf{0.5379} & - & 0.7054 & -  \\
        \midrule
        WUPSame$_{01}$ & 0.3028 & -15.95 & 0.4517 & -16.02 & 0.6065 & -14.02 \\
        UnWUPSame$_{01}$ & 0.3091 & -14.21 & 0.4655 & -13.45 & 0.6599 & -6.45 \\
        WUPUniq$_{01}$ & 0.3400 & -5.63 & 0.5000 & -7.04 & 0.6886 & -2.38 \\
        UnWUPUniq$_{01}$ & 0.3125 & -13.26 & 0.4586 & -14.74 & 0.6028 & -14.54 \\
        \midrule
        WUPSame & 0.3932$\ssymbol{1}$$\ssymbol{2}$$\ssymbol{3}$$\ssymbol{4}$$\ssymbol{5}$ & 9.13 & 0.5138$\ssymbol{1}$$\ssymbol{2}$$\ssymbol{4}$ & -4.48  & 0.6969$\ssymbol{1}$$\ssymbol{2}$ & -1.20  \\
        UnWUPSame & 0.3982$\ssymbol{1}$$\ssymbol{2}$$\ssymbol{3}$$\ssymbol{4}$$\ssymbol{5}$& 10.51 & 0.5241$\ssymbol{1}$$\ssymbol{2}$$\ssymbol{4}$ & -2.56  & 0.6952$\ssymbol{1}$$\ssymbol{2}$ & -1.44   \\
        WUPUniq & 0.3891$\ssymbol{1}$$\ssymbol{2}$$\ssymbol{3}$$\ssymbol{4}$ & 7.99 & 0.5310$\ssymbol{1}$$\ssymbol{2}$$\ssymbol{3}$$\ssymbol{4}$ & -1.28  & 0.6969$\ssymbol{1}$$\ssymbol{2}$ & -1.20   \\
        UnWUPUniq & \textbf{0.4064}$\ssymbol{1}$$\ssymbol{2}$$\ssymbol{3}$$\ssymbol{4}$$\ssymbol{5}$$\ssymbol{6}$ & 12.79 & 0.5345$\ssymbol{1}$$\ssymbol{2}$$\ssymbol{4}$ & -0.63  & \textbf{0.7106}$\ssymbol{1}$$\ssymbol{2}$$\ssymbol{3}$ & 0.73   \\
    \hline
    \end{tabular}}
\end{center}
\caption{Comparison with baseline method considering 2016 tags for training
  tag embedding. Superscript $\ssymbol{1},\ \ssymbol{2},\ \ssymbol{3},\
  \ssymbol{4},\ \ssymbol{5},$ and $\ssymbol{6}$ respectively denote the
  significant difference with UAmsterdamDL, LinearCatRev, USI5,
  DUTH\_rocchio, PK-Boosting and CS-L2Rank (computed using paired t-test
  with 95\% confidence). $\Delta$ values(\%) express the relative
  improvement compare to CS-L2Rank, the best performing baseline. Methods
  using the one-hot encoded representation of tags are indicated by
  subscript $01$ and corresponding performance are reported in second part.
  The performance of a method using dense embedding is significantly better
  than the method when one-hot encoded representation is used. For each
  experiment, reported result uses best optimization function score to find
  parameters. The best performance bold faced. No significant difference in
  performance is observed between WUPSame, UnWUPSame, WUPUniq and
  UnWUPUniq.}
\label{table:CompareWithbase}
\end{table}

\begin{table}
\begin{center}
    {\begin{tabular}{ l  l l l l l l }
  \hline
      & NDCG@5 & $\Delta(\%)$  & P@5 & $\Delta(\%)$  & MRR & $\Delta(\%)$  \\
   \hline
        UAmsterdamDL~\cite{hashemi2016neural} & 0.2824 & - & 0.4448 & - & 0.5924 & -  \\
        LinearCatRev~\cite{aliannejadiLinearCatRev} & 0.3213 & - & 0.4897 & - & 0.6284 & - \\
        USI5~\cite{aliannejadi2016venue} & 0.3265 & - & 0.5069 & - & 0.6796 & -  \\
        DUTH\_rocchio~\cite{kalamatianos2016recommending} & 0.3306 & - & 0.4724 & - & 0.6801 & - \\
        PK-Boosting~\cite{aliannejadi2018personalized} & 0.3526 & - & 0.5310 & - & 0.6800  & - \\
        CS-L2Rank~\cite{aliannejadi2017venue} & 0.3603 & - & 0.5379 & - & 0.7054 & -  \\
        \midrule
        WUPSame$_{01}$ & 0.3028 & -15.95 & 0.4517 & -16.02 & 0.6065 & -14.02 \\
        UnWUPSame$_{01}$ & 0.3091 & -14.21 & 0.4655 & -13.45 & 0.6599 & -6.45 \\
        WUPUniq$_{01}$ & 0.3400 & -5.63 & 0.5000 & -7.04 & 0.6886 & -2.38 \\
        UnWUPUniq$_{01}$ & 0.3125 & -13.26 & 0.4586 & -14.74 & 0.6028 & -14.54 \\
        \midrule
        WUPSame & \textbf{0.4067}$\ssymbol{1}$$\ssymbol{2}$$\ssymbol{3}$$\ssymbol{4}$$\ssymbol{5}$$\ssymbol{6}$ & 12.87 & 0.5586$\ssymbol{1}$$\ssymbol{2}$$\ssymbol{3}$$\ssymbol{4}$ & 3.84 & 0.7415$\ssymbol{1}$$\ssymbol{2}$$\ssymbol{3}$ & 5.11\\
        UnWUPSame & 0.4046$\ssymbol{1}$$\ssymbol{2}$$\ssymbol{3}$$\ssymbol{4}$$\ssymbol{5}$$\ssymbol{6}$ & 12.29 & \textbf{0.5621}$\ssymbol{1}$$\ssymbol{2}$$\ssymbol{3}$$\ssymbol{4}$ & 4.49 & 0.7445$\ssymbol{1}$$\ssymbol{2}$$\ssymbol{3}$ & 5.54\\
        WUPUniq & 0.3973$\ssymbol{1}$$\ssymbol{2}$$\ssymbol{3}$$\ssymbol{4}$$\ssymbol{5}$ & 10.26 & 0.5448$\ssymbol{1}$$\ssymbol{2}$$\ssymbol{4}$ & 1.28 & 0.7413$\ssymbol{1}$$\ssymbol{2}$$\ssymbol{3}$ & 5.08\\
        UnWUPUniq & 0.3796$\ssymbol{1}$$\ssymbol{2}$$\ssymbol{3}$$\ssymbol{4}$ & 5.35 & 0.5172$\ssymbol{1}$$\ssymbol{2}$ & -3.84 &\bfseries 0.7608$\ssymbol{1}$$\ssymbol{2}$$\ssymbol{3}$$\ssymbol{4}$ & 7.85\\
    \hline
    \end{tabular}}
\end{center}
\caption{Comparison with the baseline considering 2015+2016 tags for
  training tag embedding. Superscript $\ssymbol{1},\ \ssymbol{2},\
  \ssymbol{3},\ \ssymbol{4},\ \ssymbol{5},$ and $\ssymbol{6}$ respectively
  denote the significant difference with UAmsterdamDL, LinearCatRev, USI5,
  DUTH\_rocchio, PK-Boosting and CS-L2Rank (computed using paired t-test
  with 95\% confidence). $\Delta$ values (\%) express the relative
  improvement compare to CS-L2Rank, the best performing baseline. Methods
  using the one-hot encoded representation of tags are indicated by
  subscript $01$ and corresponding performance are reported in second part.
  The performance of a method using dense embedding is significantly better
  than the method when one-hot encoded representation is used. For each
  experiment, reported result uses best optimization function score to find
  parameters. The best performance bold faced. No significant difference in
  performance is observed between WUPSame, UnWUPSame, WUPUniq and
  UnWUPUniq.}
\label{table:all_emb_Method}
\end{table}

\section{Results and Discussions} \label{sec:discussion}

The performance of the proposed methods, along with the baselines are reported in Tables~\ref{table:CompareWithbase} and~\ref{table:all_emb_Method}.
In Table~\ref{table:CompareWithbase}, where TCST 2016 dataset has been employed for the training of the embedding model, the performance of the word embedding based models (i.e.  WUPSame,UnWUPSame, WUPUniq and UnWUPUniq) are observed to be almost always producing the best performance in terms of all three evaluation metrics.
The improvements over the baselines are also significant (paired t-test with 95\% confidence interval) in a majority of the cases.
Further, we notice no significant difference in performance between the proposed methods.
The performance of the methods when the combined TCST 2015 and 2016 datasets are used for training the embedding model has been reported in Table~\ref{table:all_emb_Method}.
Also, no significant differences in performance between the proposed methods using TCST 2016 dataset and combined TCST 2015-2016 datasets are observed. 
Among the baseline methods, CS-L2Rank~\cite{aliannejadi2017venue} and PK-Boosting~\cite{aliannejadi2018personalized} are seen to the best performing models.
Overall, the proposed method with weighted user profiling and same parameter settings (WUPSame) performs the best. 
Although P@5 and MRR are not significantly better than the strongest baselines for most of the cases, the proposed methods consistently achieve the best performance as compared to all the baseline methods.

The performances of the methods using one-hot encoded sparse representation of the tags are presented in the second half of Tables~\ref{table:CompareWithbase} and \ref{table:all_emb_Method}.
Note that, the performances of the four methods that use the sparse representations (specifically \textbf{WUPSame$_{01}$}, \textbf{UnWUPSame$_{01}$},
\textbf{WUPUniq$_{01}$}, and \textbf{UnWUPUniq$_{01}$}) do not require any dataset, other than the list of tags, to form the one-hot representations.
Hence, the performances reported in Tables~\ref{table:CompareWithbase} and \ref{table:all_emb_Method} are identical for these methods.
If we compare the models with their one-hot encoded alternatives, we can notice that all the models that use the dense embedded representation perform better (significantly) than their substitutes. 
This observation empirically confirms that the dense representation of the tags is indeed a valuable component of the proposed models.

\subsection{Result Analysis}
While modeling tag relationships to represent the POIs and users in a dense form, there are parameters such as the number of iterations and the size of the dataset, which control the quality of the trained model.
On top of the tag relationships, we apply different weighting techniques to model the user profiles before ranking the candidate POIs.
In sum, there are fundamentally five components that control the overall performance of the proposed models:
\begin{enumerate}[I)]
    \item \textbf{Representation of tags.} We experiment with dense, word embedding-based representation of tags to model the POIs and users. There are parameters like the number of iterations, size of the dataset to use for training etc. (see Section~\ref{subsubsec:train-embed}). 
    \item \textbf{Model parameters.} The three model parameters, $\alpha$, $\beta$, and $\gamma$ control the weights of positive, neutral, and negative feedback while constructing the user model.
    \item \textbf{Strategies to assign parameter settings while modeling users.} We study the performance of the proposed method when the identical settings are applied across all users; also, we set the parameters in a user-specific form to examine the performance (discussed in Section~\ref{subsubsec:sameparam} and~\ref{subsubsec:uniqparam}).
    \item \textbf{Weighing the user models.} We construct an unweighted model where no discrimination is adopted to distinguish ratings with strong opinions from ratings with weak opinions. Utilizing the scaled ratings, we also consider a weighted variant where strong and weak opinions are considered separately (see Section~\ref{sec:experiments}).
    \item \textbf{Metric used for performance optimization.} To avoid the possibilities of overfitting, we analyse and report the performance when different metrics are used for optimizing the parameter values (see Section~\ref{sec:parameters}).
\end{enumerate}
Changes in either of the above components would result in a deviation in the performance.
In this section, we will discuss the observations while dissecting the contribution of these components to the proposed models.
%
%



\subsubsection{Representation of tags}
We experiment with the proposed methods using tag embeddings learned with two different datasets (as discussed in Section~\ref{subsubsec:train-embed}). 
Along with the baselines, we present all our experiments performed using the tag embedding learned on the TCST 2016 dataset in Table~\ref{table:CompareWithbase}. 
In Table~\ref{table:all_emb_Method}, the result of the proposed methods, when the embedding model is trained with a combined TCST 2015 - 2016 datasets, are reported. 
%
%
%
When the TCST 2016 dataset is only utilized while training the embedding model, the unweighted model with unique parameter settings for each user (UnWUPUniq) achieves the best performance. 
We accomplish the overall best performance when the weighted model and the same parameter settings for all user profiles (WUPSame) is used on the embedding model trained with the tag information from TCST 2015 and 2016 datasets.
Although the improvements in terms of precision at rank 5 (P@5) and mean reciprocal rank (MRR) are not significant over the strongest baselines (CS-L2Rank and PK-Boosting), the proposed methods attain around 13\% improvement in terms of normalized discounted cumulative gain at rank 5 (NDCG@5) over CS-L2Rank~\cite{aliannejadi2017venue} which is a state-of-the-art method.
When the TCST 2016 tag data is only used, the achieved P@5 and MRR for all proposed methods are seen to be mostly inferior to or equivalent to PK-Boosting and CS-L2Rank, although the differences are seen to be insignificant (see Table~\ref{table:CompareWithbase}).  
As evident from Tables~\ref{table:CompareWithbase} and \ref{table:all_emb_Method}, the proposed methods perform better when both 2015 and 2016 tag information are utilized for training the embedding model.
This is an expected observation as embedding models can make better representations when the training data is relatively large.

\begin{figure}[!h]
  \centering
  \subfloat[Embedding trained on TREC CS 2015, 2016 dataset.]
  {\label{fig:ndcg_5_barchart_2015-2016}
  \includegraphics[width=0.45\textwidth]{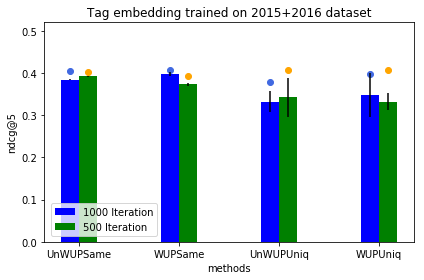}
  }
  \subfloat[Embedding trained on TREC CS 2016 dataset.] 
    {\label{fig:ndcg_5_barchart_2016}
    \includegraphics[width=0.45\textwidth]{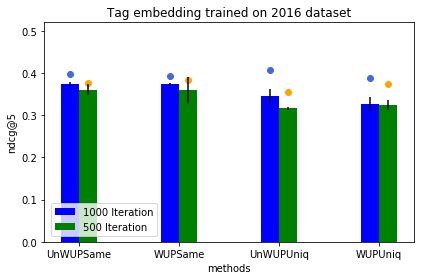}
    }
  \caption{Effect on NDCG@5 by changing the number of iterations to train the tag embedding}  \label{fig:ndcg_5_barchart}
\end{figure}

\begin{figure}[!h]
  \centering
  \subfloat[Embedding trained on TREC CS 2015, 2016 dataset.]
  {\label{fig:p_5_barchart_2015-2016}
  \includegraphics[width=0.45\textwidth]{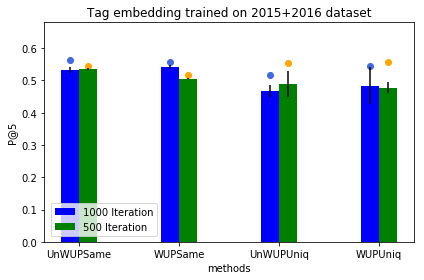}
  }
  \subfloat[Embedding trained on TREC CS 2016 dataset.] 
    {\label{fig:p_5_barchart_2016}
    \includegraphics[width=0.45\textwidth]{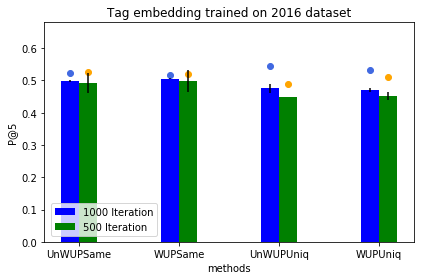}
    }
  \caption{Effect on P@5 by changing the number of iterations to train the tag embedding}  \label{fig:p_5_barchart}
\end{figure}

\begin{figure}[!h]
  \centering
  \subfloat[Embedding trained on TREC CS 2015, 2016 dataset.]
  {\label{fig:mrr_barchart_2015-2016}
  \includegraphics[width=0.45\textwidth]{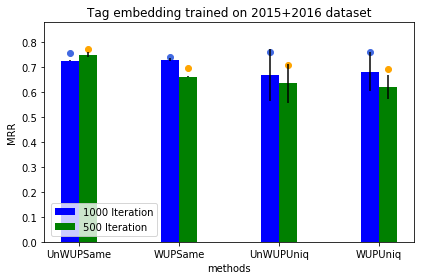}
  }
  \subfloat[Embedding trained on TREC CS 2016 dataset.] 
    {\label{fig:mrr_barchart_2016}
    \includegraphics[width=0.45\textwidth]{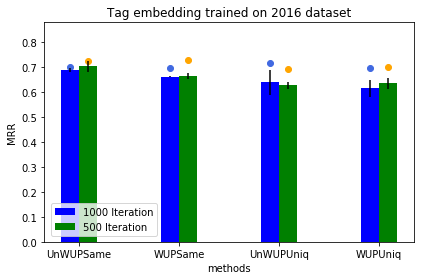}
    }
  \caption{Effect on MRR by changing the number of iterations to train the tag embedding}  \label{fig:mrr_barchart}
\end{figure}

Following the exposition reported in~\cite{diaz-etal-2016-query}, we have varied the number of iterations up to larger values to indemnify the limitation of the tag dataset with limited size.
It is evident that there is an implicit relationship between the size of the dataset and the optimal number of iterations to be applied for training the embedding model to achieve the best performance.
Hence, we experiment with 2016 as well as 2015, 2016 taken together.
The iteration parameter is varied in a dispersed range (discussed in Section~\ref{subsec:wordembed}).
To better contrast the variation in performance, we only report the results for iterations 500 and 1000, which produce the best performances among the other values.
We plot bar charts showing the average value of the performance metrics while choosing different measures to optimize the parameters ($\alpha$, $\beta$, and $\gamma$) for user profile generation.
The performance variations are graphically plotted for NDCG@5, P@5, and MRR, respectively in Figures~\ref{fig:ndcg_5_barchart},~\ref{fig:p_5_barchart}, and~\ref{fig:mrr_barchart}.
Each figure has two sub-figures that present the results on TCST 2015-2016 tag data taken together, and TCST 2016 tag data. 
The blue bars show the performance when $1000$ iterations are applied while $500$ iterations are presented with a green bar.
The maximum value is also shown with a blue and yellow point, respectively, for 1000 and 500 iterations on top of each bar.
Together with the average values, we show the standard deviations of the achieved metrics by a vertical line on the top of each bar. 

From the figures, it can be observed that changing the number of iterations affects the methods' performance.
If we consider the same parameters for all user profiles (WUPSame and UnWUPSame), the results using 1000 iteration tag embedding have a minor standard deviation. 
If we consider UnWUPUniq and WUPUniq, where the unique parameters are used per profile,
increasing the number of iterations increases the standard deviation of the results. 
An exception is observed when the additional 2015 dataset is used for trainingl; the standard deviations, in that case, are seen to be less stretched for UnWUPUniq for 1000 iterations (Figures~\ref{fig:ndcg_5_barchart_2015-2016} and \ref{fig:p_5_barchart_2015-2016}).

From the figures, we can also observe that if more data are used for training, the standard deviation of the performance increases when parameter values are trained per user (specifically UnWUPUniq and WUPUniq). 
Similar observations are noted for UnWUPUniq and WUPUniq when the iterations are increased.
An opposite phenomenon is noticed for the methods for which the same parameter values are used for all the users; increasing the dataset size and number of iterations resulted in a set of results having a lower standard deviation.
When the parameters are tuned per user, the parameters are getting overfitted for that particular user with the possibility of lack of generalization.
When the same parameters are employed for all the user profiles, the parameters are getting selected based on an extrapolated scenario.
This justifies the reason that leads to a lesser deviation when the same parameter settings are used for all the users.
Another observation from Figure~\ref{fig:ndcg_5_barchart}~-~\ref{fig:mrr_barchart} is that when the same parameters for all users are used, the maximum attained values remain close to the average values.
This shows that the score deviates less on changing the optimization measure when the same parameters are applied for all profiles.

\subsubsection{Model parameters}

\begin{figure}[!h]
  \centering
   \subfloat[Unweighted user profiling method] 
  {\label{fig:ndcg_param_opt_ablation_unw}
  \includegraphics[width=0.48\textwidth]{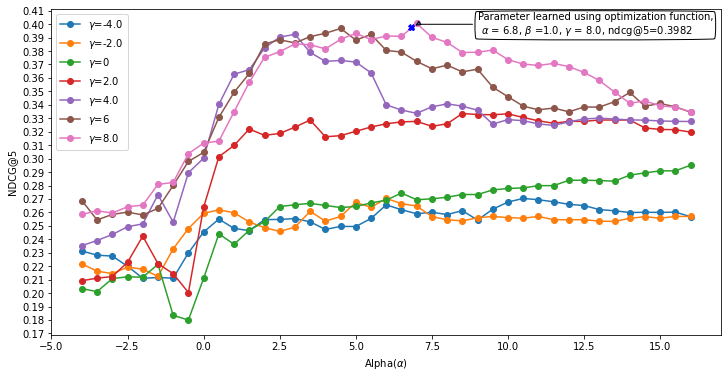}
  }
  \subfloat[Weighted user profiling method] 
    {\label{fig:ndcg_param_opt_ablation_w}
    \includegraphics[width=0.48\textwidth]{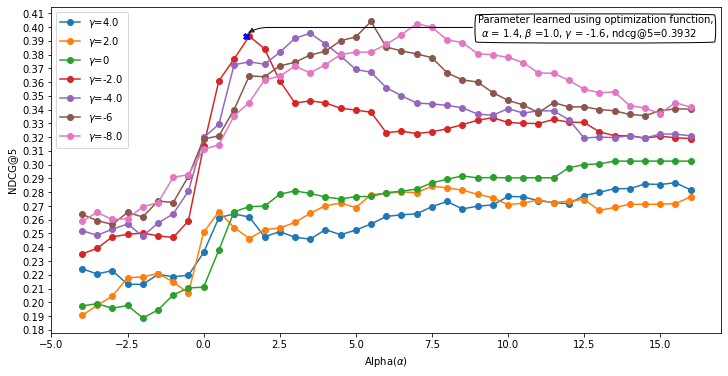}

    }
  \caption{Effect on NDCG@5 by setting different parameter values. }  \label{fig:ndcg_param_opt_ablation}
\end{figure}

\begin{figure}[!h]
  \centering
    \subfloat[Unweighted user profiling method] 
  {\label{fig:p_5_param_opt_ablation_unw}
  \includegraphics[width=0.5\textwidth]{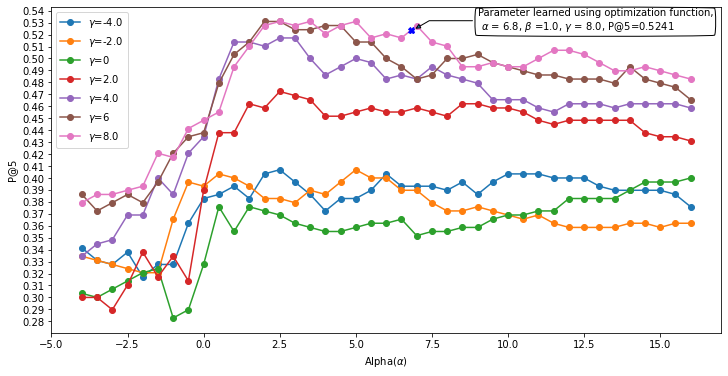}
  }
    \subfloat[Weighted user profiling method] 
    {\label{fig:p_5_param_opt_ablation_w}
    \includegraphics[width=0.48\textwidth]{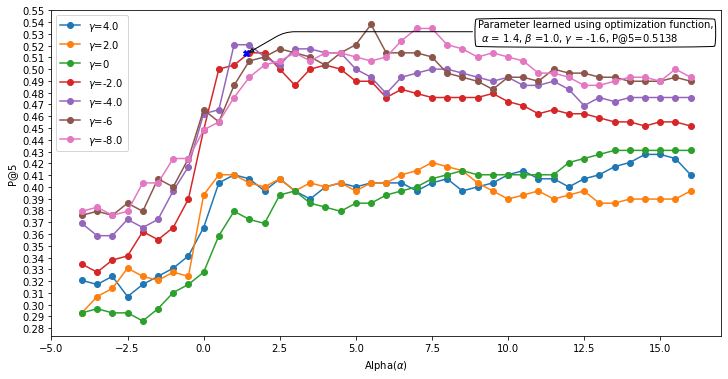}
    }
  \caption{Effect on P@5 by setting different parameter values. }  \label{fig:p_5_param_opt_ablation}
\end{figure}

\begin{figure}[!h]
  \centering
  \subfloat[Unweighted user profiling method]
  {\label{fig:mrr_param_opt_ablation_unw}
  \includegraphics[width=0.5\textwidth]{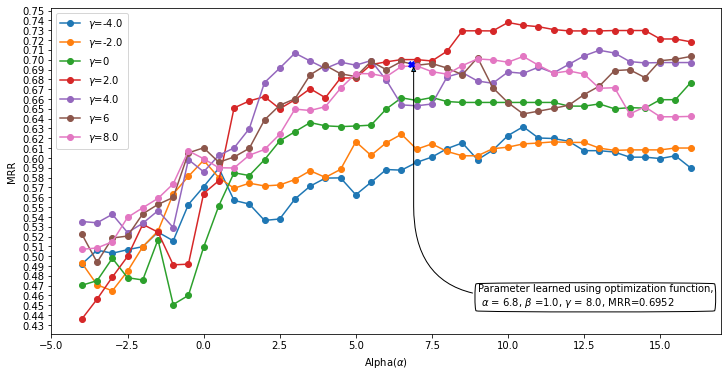}
  }
  \subfloat[Weighted user profiling method] 
    {\label{fig:mrr_param_opt_ablation_w}
    \includegraphics[width=0.48\textwidth]{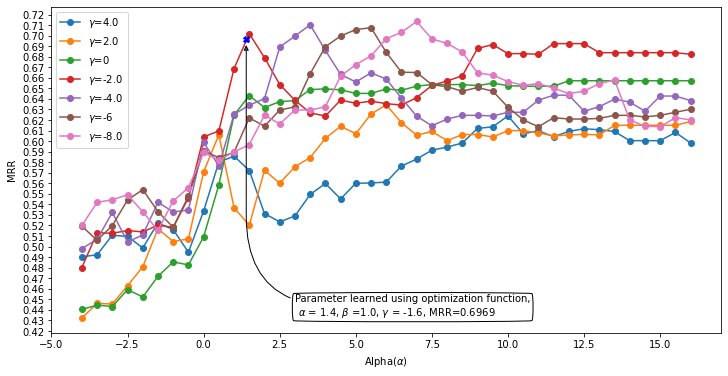}

    }
  \caption{Effect on MRR by setting different parameter values. }  \label{fig:mrr_param_opt_ablation}
\end{figure}

As discussed in Section~\ref{sec:approach}, there are three parameters associated with the model, specifically $\alpha$, $\beta$, and $\gamma$ that respectively denote the positive, neutral, and negative profiles in Equations~\ref{eq:unweighted-poi-rocchio} and~\ref{eq:weighted-poi-rocchio}.
Following the discussion in Section~\ref{sec:parameters}, the neutral weight parameter $\beta$ is set to 1.0 across all parameter settings.
To obtain the optimal value of $\alpha$ and $\gamma$, we varied them from $-8.0$ to $8.0$.
While varying $\alpha$ and $\gamma$, with the same parameters used across all user models, the deviation in performance (in terms of NDCG@5, P@5 and MRR) of the methods are presented in Figures~\ref{fig:ndcg_param_opt_ablation}, ~\ref{fig:p_5_param_opt_ablation}, and ~\ref{fig:mrr_param_opt_ablation}.
The X-axis in each figure presents the value of $\alpha$ while Y-axis shows the corresponding performance measures. 
The multiple lines in each figure correspond to the result for different settings of parameter $\gamma$. 
The performances of the methods are plotted when $\alpha$ is varied till $16.0$.
Figures~\ref{fig:ndcg_param_opt_ablation_unw},~\ref{fig:p_5_param_opt_ablation_unw} and~\ref{fig:mrr_param_opt_ablation_unw} show the performance variation of the unweighted model while the similar variation in the weighted model is presented in Figures~\ref{fig:ndcg_param_opt_ablation_w},~\ref{fig:p_5_param_opt_ablation_w} and~\ref{fig:mrr_param_opt_ablation_w}.
The attained performance with the parameter values determined using the optimization method (described in Section~\ref{sec:parameters}) is also indicated using a blue cross ($\mathbf{\textcolor{blue}{x}}$). 
From the figures, it can be observed that the blue crosses are close to the peak of the plots. 
This indicates that the parameter optimization technique using the \term{profile} information, described in Section~\ref{sec:parameters}, provides a legitimate approximation.
Note that, the performance variation of the queries are reported in these figures, but the parameter tuning is performed solely utilizing the \term{profiles} and the queries (along with the relevance judgements) were not used while tuning.

\subsubsection{Strategies to assign parameter values while modeling users}

\begin{figure}[!h]
  \centering
    \subfloat[Unweighted user profile] 
  {\label{fig:unw_prec_recall_param}
  \includegraphics[width=0.45\textwidth]{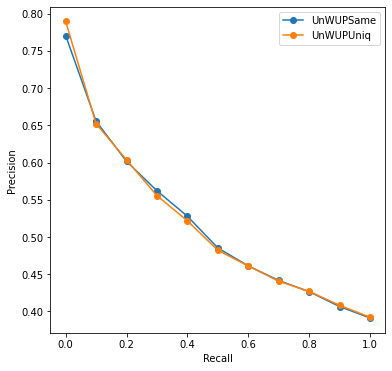}
  }
    \subfloat[Weighted user profile] 
    {\label{fig:w_prec_recall_param}
    \includegraphics[width=0.45\textwidth]{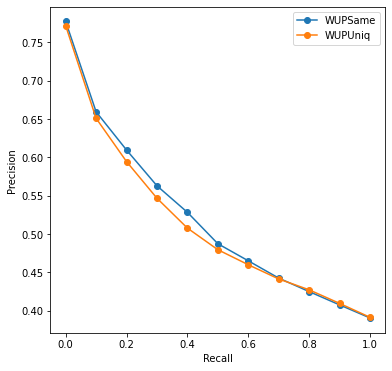}
    }
  \caption{11-point interpolated precision-recall curve for different weighing of user profiles (same parameter across all users versus unique parameters per user).}  \label{fig:2016_precision_recall_parameter}
\end{figure}

To determine the effect of applying the optimal parameter settings, either the same parameter across all users, or unique parameters per user (defined in Section~\ref{sec:parameters}), we plot 11-point precision-recall graph across all queries in Figure~\ref{fig:2016_precision_recall_parameter}. 
In the plot, we have shown the interpolated precision values over all queries at different recall levels. 
The two sub-figures correspond to the unweighted and weighted user models.
The orange line plot shows the average interpolated precision-recall curve when unique parameter settings are used, while the blue curve displays the method with the same parameter settings for all users. 
The methods with a unique parameter per user settings (WUPUniq, and UnWUPUniq) do not have a significant difference with same parameter per-user models (WUPSame, and UnWUPSame), but it is evident from the plot that applying the same parameter settings is mostly performing better than assigning unique parameters per user.

\subsubsection{Weighing the user models}

\begin{figure}[!h]
  \centering
    \subfloat[Same Rochhio parameter across different users] 
  {\label{fig:same_prec_recall_profile}
  \includegraphics[width=0.45\textwidth]{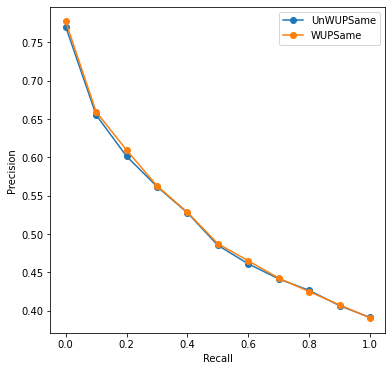}
  }
    \subfloat[Unique Rochhio parameter across different users] 
    {\label{fig:unique_prec_recall_profile}
    \includegraphics[width=0.45\textwidth]{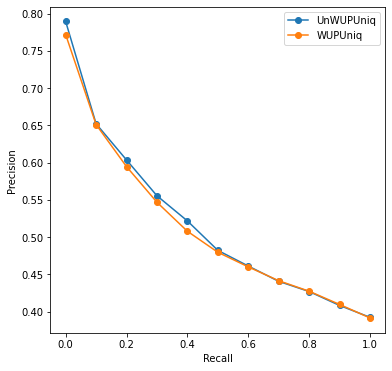}
    }
  \caption{11-point interpolated precision-recall curve for different user profile settings (weighted versus unweighted).}  \label{fig:2016_precision_recall_user_prof}
\end{figure}

A similar graph is also presented in Figure~\ref{fig:2016_precision_recall_user_prof} to observe the effect of considering the weights while modeling the user.
Each figure shows the comparison between the optimized weighted user model with the unweighted counterpart using line plots. 
The orange line depicts the weighted user model, while the blue one shows the interpolated precision-recall curve when the unweighted user modeling is used. 
In Figure~\ref{fig:same_prec_recall_profile}, the interpolated precision of the methods with weighted and unweighted user models (but applying the same settings for all user models) is presented. 
Noticeably, there is barely any difference in performance as observed from the figure which indicates that the discrete parameter settings is selecting the optimal values for which the methods are producing similar performances.
When user-specific parameters are applied (UnWUPUniq, and WUPUniq), the performance variation is shown in Figure~\ref{fig:unique_prec_recall_profile} that shows slightly better performance of the unweighted model in terms of early precision.
However, if we observe the line chart in Figures~\ref{fig:ndcg_param_opt_ablation},~\ref{fig:p_5_param_opt_ablation} and~\ref{fig:mrr_param_opt_ablation}, the optimal value of parameters (indicated by the blue cross) is attaining for smaller values of $\alpha$, and $\gamma$ for the weighted variant ($\alpha = 1.4, \gamma = -1.6$).
In comparison, the optimal values for the unweighted model are obtained with $\alpha = 6.8$ and $\gamma = -8.0$.
The reason for this is due to the scaling of profile vectors by the weights in the weighted profiling, which is being compensated by larger parameter values in unweighted user modeling.

\subsubsection{Metric used for performance optimization}
\label{subsec:param-tune}

The parameters of the proposed methods ($\alpha$, $\beta$, and $\gamma$) are tuned based on optimizing the performance in terms of NDCG@5, and the corresponding results are reported in Tables~\ref{table:CompareWithbase} and \ref{table:all_emb_Method}.
To validate the robustness of the proposed methods by confirming that the performances are not the results of overfitting on certain parameter values (in this case NDCG@5), we experiment on how the performances (in terms of P@5, NDCG@5, and MRR) vary when the optimization measure, dataset and the number of iterations are changed.
Specifically, we report the performances by changing optimization measures (to average precision, recall, NDCG at different ranks, reciprocal rank, bpref, etc.) to optimize the parameters, tag datasets used for training the embedding model (only 2016 tags or, considering 2015 along with 2016 tag information) and, number of iteration (500 and 1000) used while training the tag embedding across different methods.
To show the performance variation, we plot the heat maps with the varying parameters in Tables~\ref{table:Heatmap_ndcg@5_2015+2016}~-~\ref{table:Heatmap_MRR_2016}. 
The first column in the table indicates the optimization measure that has been used to tune the parameters ($\alpha$, $\beta$ and $\gamma$) and the corresponding row contains the achieved performance measured in terms of the respective metrics.
Two major columns in all the tables represent the number of iterations (respectively 500 and 1000) used to train the model to get tag embeddings; the four sub-columns in each major column present the performance of UnWUPSame, WUPSame, UnWUPUniq, WUPUniq respectively. 
We report the results in terms of NDCG@5, P@5 and MRR respectively in Table~\ref{table:Heatmap_ndcg@5_2015+2016}~-~\ref{table:Heatmap_ndcg@5_2016}, Table~\ref{table:Heatmap_P@5_2015+2016}~-~\ref{table:Heatmap_P@5_2016} and Table~\ref{table:Heatmap_MRR_2015+2016}~-~\ref{table:Heatmap_MRR_2016}.
We first report the results for each measure when both 2015 and 2016 tag information are used for training (Tables~\ref{table:Heatmap_ndcg@5_2015+2016}, \ref{table:Heatmap_P@5_2015+2016} and \ref{table:Heatmap_MRR_2015+2016}) followed by the results when only 2016 tag are utilized (in Tables~\ref{table:Heatmap_ndcg@5_2016}, \ref{table:Heatmap_P@5_2016} and \ref{table:Heatmap_MRR_2016}).

It is observed from the heat maps that the rows for average precision (AP) and normalized discounted cumulative gain (NDCG) are on an average darker in comparison to other metrics; this indicates that AP and NDCG are in general good optimization metrics based on which the parameters can be trained.
These measures (AP and NDCG) depend on the top ranked relevant POIs, which is appropriate for our recommendation scenario, and hence, they work better then the others for training the parameters.
It can also be noticed that the recall (specifically recall at rank 20) is also a good optimizing metric when the 2015 and 2016 tag information are used together, as seen in Tables~\ref{table:Heatmap_ndcg@5_2015+2016},~\ref{table:Heatmap_P@5_2015+2016} and~\ref{table:Heatmap_MRR_2015+2016}.
However, the optimal performances for other settings are not attainable using parameters trained on recall.

If we compare the performance when TCST 2015 and 2016 datasets are utilized together (Tables~\ref{table:Heatmap_ndcg@5_2015+2016},~\ref{table:Heatmap_P@5_2015+2016}, and~\ref{table:Heatmap_MRR_2015+2016}) with the results when only 2016 tag data is used (Tables~\ref{table:Heatmap_ndcg@5_2016},~\ref{table:Heatmap_P@5_2016}, and~\ref{table:Heatmap_MRR_2016}), it can be noted that the rows are comparatively darker in the first set of tables. 
This indicates that the methods achieve better performance when 2015 and 2016 datasets are used together.
As discussed earlier in this section, this verifies that adding more tag data corresponding to POIs can improve the result further.

\pgfplotstableset{
    /color cells/min/.initial=0,
    /color cells/max/.initial=1000,
    /color cells/textcolor/.initial=,
    %
    color cells/.code={%
        \pgfqkeys{/color cells}{#1}%
        \pgfkeysalso{%
            postproc cell content/.code={%
                \begingroup
                %
                \pgfkeysgetvalue{/pgfplots/table/@preprocessed cell content}\value
\ifx\value\empty
\endgroup
\else
                \pgfmathfloatparsenumber{\value}%
                \pgfmathfloattofixed{\pgfmathresult}%
                \let\value=\pgfmathresult
                %
                \pgfplotscolormapaccess
                    [\pgfkeysvalueof{/color cells/min}:\pgfkeysvalueof{/color cells/max}]%
                    {\value}%
                    {\pgfkeysvalueof{/pgfplots/colormap name}}%
                %
                \pgfkeysgetvalue{/pgfplots/table/@cell content}\typesetvalue
                \pgfkeysgetvalue{/color cells/textcolor}\textcolorvalue
                %
                \toks0=\expandafter{\typesetvalue}%
                \xdef\temp{%
                    \noexpand\pgfkeysalso{%
                        @cell content={%
                            \noexpand\cellcolor[rgb]{\pgfmathresult}%
                            \noexpand\definecolor{mapped color}{rgb}{\pgfmathresult}%
                            \ifx\textcolorvalue\empty
                            \else
                                \noexpand\color{\textcolorvalue}%
                            \fi
                            \the\toks0 %
                        }%
                    }%
                }%
                \endgroup
                \temp
\fi
            }%
        }%
    }
}

\begin{table}
\vrule
\centering
\tiny
\pgfplotstabletypeset[
    every head row/.style={%
        before row={\toprule
                &\multicolumn{4}{c}{Embedding Trained on 2015+2016 Tags 500 Iteration} & \multicolumn{4}{c}{Embedding Trained on 2015+2016 Tags 1000 Iteration}\\
                \midrule
            },
        after row=\midrule
    },
    every last row/.style={after row=\bottomrule},
    color cells={min=0.21,max=0.6,textcolor=black},
    /pgfplots/colormap = {blackwhite}{gray(0cm)=(1); gray(1cm)=(0)},
    /pgf/number format/fixed,
    /pgf/number format/fixed zerofill,
    /pgf/number format/precision=4,
    col sep=comma,
    columns/NDCG@5/.style={reset styles,string type},
        columns/UnWUPSame/.style={
             column name={$UnWUPSame$},
        },
        columns/WUPSame/.style={
              column name={$WUPSame$},
        },
        columns/UnWUPUniq/.style={
             column name={$UnWUPUniq$},
        },
        columns/WUPUniq/.style={
              column name={$WUPUniq$},
        },
        columns/UnWUPSame1/.style={
             column name={$UnWUPSame$},
        },
        columns/WUPSame1/.style={
              column name={$WUPSame$},
        },
        columns/UnWUPUniq1/.style={
             column name={$UnWUPUniq$},
        },
        columns/WUPUniq1/.style={
              column name={$WUPUniq$},
        },
]{%
NDCG@5,UnWUPSame,WUPSame,UnWUPUniq,WUPUniq,UnWUPSame1,WUPSame1,UnWUPUniq1,WUPUniq1
bpref,0.3973,0.3462,0.3474,0.3410,0.3936,0.3918,0.3171,0.3490
ap,0.3973,0.3668,0.3598,0.3694,0.3954,0.4067,0.3683,0.3722
ap\_at\_5,0.4023,0.3813,0.3142,0.3288,0.3898,0.3921,0.3377,0.3381
ap\_at\_10,0.4006,0.3813,0.3786,0.3678,0.3866,0.4009,0.3796,0.3769
ap\_at\_20,0.3973,0.3668,0.4077,0.4061,0.3882,0.3915,0.3613,0.3796
NDCG,0.4008,0.3888,0.3379,0.3364,0.4013,0.4067,0.3592,0.3545
NDCG\_at\_5,0.3995,0.3861,0.3267,0.3535,0.3968,0.4067,0.3462,0.3614
NDCG\_at\_10,0.4006,0.3888,0.3515,0.3325,0.3975,0.4067,0.3286,0.3533
NDCG\_cut\_20,0.4004,0.3668,0.3455,0.3210,0.3996,0.4067,0.3350,0.3973
P\_5,0.3942,0.3839,0.3136,0.3063,0.4046,0.3921,0.3075,0.3368
P\_10,0.3847,0.3822,0.3812,0.3579,0.3947,0.4068,0.3766,0.3711
P\_20,0.3977,0.3429,0.3596,0.2727,0.3873,0.4027,0.2890,0.3731
recall\_5,0.3953,0.3649,0.3136,0.3063,0.3607,0.3921,0.3075,0.3368
recall\_10,0.4012,0.3822,0.3812,0.3579,0.3947,0.4068,0.3766,0.3711
recall\_20,0.3977,0.3671,0.3596,0.2727,0.3873,0.4068,0.2890,0.3731
recip\_rank,0.3938,0.3437,0.2307,0.2464,0.3532,0.3984,0.2369,0.2135
}
\vrule
\caption{Effect on NDCG@5 on varying the measures to tune the parameters considering 2015+2016 dataset}
\label{table:Heatmap_ndcg@5_2015+2016}
\end{table}

\begin{table}
\vrule
\centering
\tiny
\pgfplotstabletypeset[
    every head row/.style={%
        before row={\toprule
                &\multicolumn{4}{c}{Embedding Trained on 2016 Tags 500 Iteration} & \multicolumn{4}{c}{Embedding Trained on 2016 Tags 1000 Iteration}\\
                \midrule
            },
        after row=\midrule
    },
    every last row/.style={after row=\bottomrule},
    color cells={min=0.21,max=0.6,textcolor=black},
    /pgfplots/colormap = {blackwhite}{gray(0cm)=(1); gray(1cm)=(0)},
    /pgf/number format/fixed,
    /pgf/number format/fixed zerofill,
    /pgf/number format/precision=4,
    col sep=comma,
    columns/NDCG@5/.style={reset styles,string type},
        columns/UnWUPSame/.style={
             column name={$UnWUPSame$},
        },
        columns/WUPSame/.style={
              column name={$WUPSame$},
        },
        columns/UnWUPUniq/.style={
             column name={$UnWUPUniq$},
        },
        columns/WUPUniq/.style={
              column name={$WUPUniq$},
        },
        columns/UnWUPSame1/.style={
             column name={$UnWUPSame$},
        },
        columns/WUPSame1/.style={
              column name={$WUPSame$},
        },
        columns/UnWUPUniq1/.style={
             column name={$UnWUPUniq$},
        },
        columns/WUPUniq1/.style={
              column name={$WUPUniq$},
        },
]{%
NDCG@5,UnWUPSame,WUPSame,UnWUPUniq,WUPUniq,UnWUPSame1,WUPSame1,UnWUPUniq1,WUPUniq1
bpref,0.3751,0.3642,0.2963,0.3189,0.3587,0.3462,0.3231,0.3164
ap,0.3666,0.3637,0.3433,0.3732,0.3783,0.3668,0.3700,0.3443
ap\_at\_5,0.3655,0.3596,0.2916,0.3308,0.3946,0.3813,0.2891,0.2864
ap\_at\_10,0.3666,0.3637,0.3071,0.3615,0.3887,0.3813,0.3572,0.3412
ap\_at\_20,0.3740,0.3604,0.3521,0.3460,0.3783,0.3668,0.3667,0.3472
NDCG,0.3604,0.3637,0.3551,0.3525,0.3924,0.3888,0.4064,0.3891
NDCG\_at\_5,0.3523,0.3622,0.3291,0.3495,0.3982,0.3861,0.3709,0.3445
NDCG\_at\_10,0.3505,0.3637,0.3514,0.3599,0.3909,0.3888,0.4008,0.3656
NDCG\_at\_20,0.3591,0.3637,0.3418,0.3614,0.3935,0.3668,0.3850,0.3646
P\_5,0.3393,0.3593,0.2907,0.3117,0.3890,0.3839,0.2862,0.2811
P\_10,0.3499,0.3622,0.2970,0.3081,0.3749,0.3822,0.3577,0.3014
P\_20,0.3691,0.3777,0.3471,0.2955,0.3504,0.3429,0.3404,0.3366
recall\_5,0.3393,0.3840,0.2907,0.3117,0.3890,0.3649,0.2862,0.2811
recall\_10,0.3499,0.3622,0.2970,0.3081,0.3749,0.3822,0.3577,0.3014
recall\_20,0.3654,0.3308,0.3471,0.2955,0.3504,0.3671,0.3404,0.3366
recip\_rank,0.3604,0.3507,0.2662,0.2418,0.3214,0.3437,0.3048,0.2181
}
\vrule
\caption{Effect on NDCG@5 on varying the measures to tune the parameters considering 2016 dataset}
\label{table:Heatmap_ndcg@5_2016}
\end{table}

\begin{table}
\vrule
\centering
\tiny
\pgfplotstabletypeset[
    every head row/.style={%
        before row={\toprule
                &\multicolumn{4}{c}{Embedding Trained on 2015+2016 Tags 500 Iteration} & \multicolumn{4}{c}{Embedding Trained on 2015+2016 Tags 1000 Iteration}\\
                \midrule
            },
        after row=\midrule
    },
    every last row/.style={after row=\bottomrule},
    color cells={min=0.32,max=0.75,textcolor=black},
    /pgfplots/colormap = {blackwhite}{gray(0cm)=(1); gray(1cm)=(0)},
    /pgf/number format/fixed,    
    /pgf/number format/fixed zerofill,
    /pgf/number format/precision=4,
    col sep=comma,
    columns/P@5/.style={reset styles,string type},
        columns/UnWUPSame/.style={
             column name={$UnWUPSame$},
        },
        columns/WUPSame/.style={
              column name={$WUPSame$},
        },
        columns/UnWUPUniq/.style={
             column name={$UnWUPUniq$},
        },
        columns/WUPUniq/.style={
              column name={$WUPUniq$},
        },
        columns/UnWUPSame1/.style={
             column name={$UnWUPSame$},
        },
        columns/WUPSame1/.style={
              column name={$WUPSame$},
        },
        columns/UnWUPUniq1/.style={
             column name={$UnWUPUniq$},
        },
        columns/WUPUniq1/.style={
              column name={$WUPUniq$},
        },
]{%
P@5,UnWUPSame,WUPSame,UnWUPUniq,WUPUniq,UnWUPSame1,WUPSame1,UnWUPUniq1,WUPUniq1
bpref,0.5379,0.4966,0.4828,0.4759,0.5345,0.5276,0.4483,0.4862
ap,0.5379,0.4966,0.5172,0.5207,0.5379,0.5586,0.5034,0.5069
ap\_at\_5,0.5448,0.5138,0.4448,0.4690,0.5310,0.5241,0.4690,0.4759
ap\_at\_10,0.5414,0.5138,0.5276,0.5172,0.5241,0.5552,0.5172,0.5172
ap\_at\_20,0.5379,0.4966,0.5552,0.5586,0.5310,0.5310,0.5000,0.5172
NDCG,0.5414,0.5172,0.4793,0.4966,0.5586,0.5586,0.5103,0.5069
NDCG\_at\_5,0.5448,0.5103,0.4931,0.5138,0.5517,0.5586,0.4828,0.5103
NDCG\_at\_10,0.5414,0.5172,0.5069,0.4862,0.5552,0.5586,0.4690,0.5034
NDCG\_at\_20,0.5414,0.4966,0.4931,0.4621,0.5552,0.5586,0.4724,0.5448
P\_5,0.5345,0.5172,0.4414,0.4448,0.5621,0.5241,0.4483,0.4759
P\_10,0.5241,0.5172,0.5483,0.5138,0.5517,0.5586,0.5103,0.5138
P\_20,0.5379,0.4931,0.5069,0.4103,0.5310,0.5483,0.4241,0.5207
recall\_5,0.5379,0.4931,0.4414,0.4448,0.5034,0.5241,0.4483,0.4759
recall\_10,0.5448,0.5172,0.5483,0.5138,0.5517,0.5586,0.5103,0.5138
recall\_20,0.5379,0.4966,0.5069,0.4103,0.5310,0.5586,0.4241,0.5207
recip\_rank,0.5379,0.4966,0.3517,0.3759,0.5000,0.5483,0.3552,0.3276
}
\vrule
\caption{Effect on P@5 on varying the measures to tune the parameters considering 2015+2016 dataset}
\label{table:Heatmap_P@5_2015+2016}
\end{table}

\begin{table}
\vrule
\centering
\tiny
\pgfplotstabletypeset[
    every head row/.style={%
        before row={\toprule
                &\multicolumn{4}{c}{Embedding Trained on 2016 Tags, 500 Iteration} & \multicolumn{4}{c}{Embedding Trained on 2016 Tags 1000 Iteration}\\
                \midrule
            },
        after row=\midrule
    },
    every last row/.style={after row=\bottomrule},
    color cells={min=0.32,max=0.75,textcolor=black},
    /pgfplots/colormap = {blackwhite}{gray(0cm)=(1); gray(1cm)=(0)},
    /pgf/number format/fixed,
    /pgf/number format/fixed zerofill,
    /pgf/number format/precision=4,
    fixed zerofill,
    col sep=comma,
    columns/P@5/.style={reset styles,string type},
        columns/UnWUPSame/.style={
             column name={$UnWUPSame$},
        },
        columns/WUPSame/.style={
              column name={$WUPSame$},
        },
        columns/UnWUPUniq/.style={
             column name={$UnWUPUniq$},
        },
        columns/WUPUniq/.style={
              column name={$WUPUniq$},
        },
        columns/UnWUPSame1/.style={
             column name={$UnWUPSame$},
        },
        columns/WUPSame1/.style={
              column name={$WUPSame$},
        },
        columns/UnWUPUniq1/.style={
             column name={$UnWUPUniq$},
        },
        columns/WUPUniq1/.style={
              column name={$WUPUniq$},
        },
]{%
P@5,UnWUPSame,WUPSame,UnWUPUniq,WUPUniq,UnWUPSame1,WUPSame1,UnWUPUniq1,WUPUniq1
bpref,0.5276,0.5138,0.4310,0.4621,0.4966,0.4966,0.4621,0.4862
ap,0.5000,0.5034,0.4690,0.5034,0.4931,0.4966,0.4897,0.4862
ap\_at\_5,0.4897,0.4897,0.4103,0.4448,0.5207,0.5138,0.4241,0.4517
ap\_at\_10,0.5000,0.5034,0.4448,0.4931,0.5103,0.5138,0.4828,0.4897
ap\_at\_20,0.5138,0.4897,0.4828,0.4828,0.4931,0.4966,0.4931,0.4966
NDCG,0.4828,0.5034,0.4793,0.4759,0.5172,0.5172,0.5345,0.5310
NDCG\_at\_5,0.4759,0.5034,0.4310,0.4552,0.5241,0.5103,0.4828,0.4793
NDCG\_at\_10,0.4759,0.5034,0.4793,0.4862,0.5138,0.5172,0.5310,0.5138
NDCG\_at\_20,0.4793,0.5034,0.4793,0.4966,0.5172,0.4966,0.5379,0.5138
P\_5,0.4621,0.4897,0.4241,0.4448,0.5138,0.5172,0.4207,0.4483
P\_10,0.4724,0.5034,0.4414,0.4448,0.4897,0.5172,0.4897,0.4414
P\_20,0.5207,0.5138,0.4897,0.4379,0.4897,0.4931,0.4690,0.4690
recall\_5,0.4621,0.5207,0.4241,0.4448,0.5138,0.4931,0.4207,0.4483
recall\_10,0.4724,0.5034,0.4414,0.4448,0.4897,0.5172,0.4897,0.4414
recall\_20,0.5103,0.4897,0.4897,0.4379,0.4897,0.4966,0.4690,0.4690
recip\_rank,0.4828,0.4828,0.4000,0.3690,0.4552,0.4966,0.4414,0.3448
}
\vrule
\caption{Effect on P@5 on varying the measures to tune the parameters considering 2016 dataset}
\label{table:Heatmap_P@5_2016}
\end{table}

\begin{table}
\vrule
\centering
\tiny
\pgfplotstabletypeset[
    every head row/.style={%
        before row={\toprule
                &\multicolumn{4}{c}{Embedding Trained on 2015+2016 Tags 500 Iteration} & \multicolumn{4}{c}{Embedding Trained on 2015+2016 Tags 1000 Iteration}\\
                \midrule
            },
        after row=\midrule
    },
    every last row/.style={after row=\bottomrule},
    color cells={min=0.46,max=1.1,textcolor=black},
    /pgfplots/colormap = {blackwhite}{gray(0cm)=(1); gray(1cm)=(0)},
    /pgf/number format/fixed,
    /pgf/number format/fixed zerofill,
    /pgf/number format/precision=4,
    col sep=comma,
    columns/MRR/.style={reset styles,string type},
        columns/UnWUPSame/.style={
             column name={$UnWUPSame$},
        },
        columns/WUPSame/.style={
              column name={$WUPSame$},
        },
        columns/UnWUPUniq/.style={
             column name={$UnWUPUniq$},
        },
        columns/WUPUniq/.style={
              column name={$WUPUniq$},
        },
        columns/UnWUPSame1/.style={
             column name={$UnWUPSame$},
        },
        columns/WUPSame1/.style={
              column name={$WUPSame$},
        },
        columns/UnWUPUniq1/.style={
             column name={$UnWUPUniq$},
        },
        columns/WUPUniq1/.style={
              column name={$WUPUniq$},
        },
]{%
MRR,UnWUPSame,WUPSame,UnWUPUniq,WUPUniq,UnWUPSame1,WUPSame1,UnWUPUniq1,WUPUniq1
bpref,0.7457,0.6163,0.6887,0.6866,0.7311,0.7067,0.6991,0.7093
ap,0.7457,0.6515,0.6370,0.6419,0.7357,0.7415,0.7397,0.6948
ap\_at\_5,0.7723,0.6655,0.5618,0.5872,0.7363,0.7167,0.6646,0.6500
ap\_at\_10,0.7440,0.6655,0.6441,0.6471,0.7553,0.7357,0.7608,0.7306
ap\_at\_20,0.7457,0.6515,0.7084,0.6772,0.7429,0.7068,0.7253,0.7311
NDCG,0.7723,0.6843,0.6518,0.6166,0.7448,0.7415,0.7097,0.7187
NDCG\_at\_5,0.7295,0.6859,0.6007,0.6653,0.7264,0.7415,0.7022,0.7198
NDCG\_at\_10,0.7440,0.6843,0.6747,0.6212,0.7318,0.7415,0.7050,0.7425
NDCG\_at\_20,0.7723,0.6515,0.6609,0.6005,0.7319,0.7415,0.7045,0.7413
P\_5,0.7384,0.6670,0.5627,0.5604,0.7445,0.7167,0.6113,0.6753
P\_10,0.7258,0.6669,0.6641,0.6230,0.7318,0.7386,0.7119,0.7011
P\_20,0.7486,0.6288,0.6719,0.5918,0.7340,0.7420,0.6319,0.6929
recall\_5,0.7723,0.6592,0.5627,0.5604,0.7244,0.7167,0.6113,0.6753
recall\_10,0.7254,0.6669,0.6641,0.6230,0.7318,0.7386,0.7119,0.7011
recall\_20,0.7486,0.6564,0.6719,0.5918,0.7340,0.7386,0.6319,0.6929
recip\_rank,0.7342,0.6231,0.5336,0.5143,0.7195,0.7288,0.4900,0.4839
}
\vrule
\caption{Effect on MRR on varying the measures to tune the parameters considering 2015+2016 dataset}
\label{table:Heatmap_MRR_2015+2016}
\end{table}

\begin{table}
\vrule
\centering
\tiny
\pgfplotstabletypeset[
    every head row/.style={%
        before row={\toprule
                &\multicolumn{4}{c}{Embedding Trained on 2016 Tags 500 Iteration} & \multicolumn{4}{c}{Embedding Trained on 2016 Tags 1000 Iteration}\\
                \midrule
            },
        after row=\midrule
    },
    every last row/.style={after row=\bottomrule},
    color cells={min=0.46,max=1.1,textcolor=black},
    /pgfplots/colormap = {blackwhite}{gray(0cm)=(1); gray(1cm)=(0)},
    /pgf/number format/fixed,
    /pgf/number format/fixed zerofill,
    /pgf/number format/precision=4,
    col sep=comma,
    columns/MRR/.style={reset styles,string type},
        columns/UnWUPSame/.style={
             column name={$UnWUPSame$},
        },
        columns/WUPSame/.style={
              column name={$WUPSame$},
        },
        columns/UnWUPUniq/.style={
             column name={$UnWUPUniq$},
        },
        columns/WUPUniq/.style={
              column name={$WUPUniq$},
        },
        columns/UnWUPSame1/.style={
             column name={$UnWUPSame$},
        },
        columns/WUPSame1/.style={
              column name={$WUPSame$},
        },
        columns/UnWUPUniq1/.style={
             column name={$UnWUPUniq$},
        },
        columns/WUPUniq1/.style={
              column name={$WUPUniq$},
        },
]{%
MRR,UnWUPSame,WUPSame,UnWUPUniq,WUPUniq,UnWUPSame1,WUPSame1,UnWUPUniq1,WUPUniq1
bpref,0.6899,0.6959,0.6389,0.6309,0.6892,0.6163,0.5899,0.5935
ap,0.7025,0.6516,0.6618,0.6776,0.6994,0.6515,0.6595,0.6349
ap\_at\_5,0.7254,0.6741,0.6194,0.6818,0.6880,0.6655,0.5970,0.5898
ap\_at\_10,0.7025,0.6516,0.6539,0.6644,0.6878,0.6655,0.6398,0.5846
ap\_at\_20,0.6996,0.6735,0.6589,0.6221,0.6994,0.6515,0.6755,0.6150
NDCG,0.7084,0.6516,0.6939,0.6735,0.6878,0.6843,0.7106,0.6969
NDCG\_at\_5,0.6999,0.6515,0.6477,0.7022,0.6952,0.6859,0.6878,0.6727
NDCG\_at\_10,0.6995,0.6516,0.6771,0.6923,0.6849,0.6843,0.7160,0.6578
NDCG\_at\_20,0.7153,0.6516,0.6242,0.6552,0.6878,0.6515,0.6721,0.6856
P\_5,0.6976,0.6649,0.6164,0.6440,0.6856,0.6670,0.5970,0.5915
P\_10,0.6949,0.6515,0.6112,0.6135,0.6943,0.6669,0.6339,0.5497
P\_20,0.7134,0.7295,0.6240,0.6020,0.6882,0.6288,0.6399,0.6258
recall\_5,0.6976,0.7060,0.6164,0.6440,0.6856,0.6592,0.5970,0.5915
recall\_10,0.6949,0.6515,0.6112,0.6135,0.6943,0.6669,0.6339,0.5497
recall\_20,0.6782,0.6694,0.6240,0.6020,0.6882,0.6564,0.6399,0.6258
recip\_rank,0.7084,0.6472,0.5321,0.5512,0.6800,0.6231,0.5442,0.4618
}
\vrule
\caption{Effect on MRR on varying the measures to tune the parameters considering 2016 dataset}
\label{table:Heatmap_MRR_2016}
\end{table}

\subsection{Privacy issue}
There are privacy concerns in mobile recommender systems as significant volumes of sensitive (personal preference) data are stored on mobile devices that can be compromised while transmitting to a remote server by some application through a \emph{man-in-the-middle attack}. 
As discussed by Arampatzis et al.~\cite{arampatzis2018suggesting}, this is a general concern for application users with respect to identity and privacy in terms of location.

The proposed method in this article can protect user privacy by locally keeping the sensitive user information. 
With the rapid advancement of technology, mobile devices nowadays are computationally intensive with significant storage capacity. 
It is possible to create user-specific profile vectors on the device by storing the preferences of previous visits without the risk of revealing any personal information. 
The user modeling based on Equation~\ref{eq:unweighted-poi-rocchio} or~\ref{eq:weighted-poi-rocchio} can be done locally based on the user's personal preferences.
The only information to be communicated from the user's end is the context or the location of the user at which the recommendation is wanted.
Based on the context information (e.g., location/city), the recommender system would suggest a list of places along with their tags and vector representations.
The tag representation and parameter settings ($\alpha$, $\beta$, and $\gamma$ of Equation~\ref{eq:unweighted-poi-rocchio} or ~\ref{eq:weighted-poi-rocchio}) can be passed to the mobile device from the server.
Using this information, the ranking can be performed locally on the device without the risk of a privacy breach. 

\section{Conclusions and Future Work} \label{sec:conclusion}
%
Recommendation of interesting places to visit is an apparent necessity for travelers. 
Other than traditional recommender system-based approaches, previous works on POI suggestion systems utilize user comments / reviews, preference keywords, and category of location for recommending potential venues to users.
Furthermore, user-assigned tags for POIs are also employed, but the tags were considered independent to each other in earlier works.
The recent development of deep learning and neural networks opens up a plethora of research opportunities in the field of text processing.
Embedding techniques have been seen to capture the semantics relationships between terms.
In this article, we have empirically shown that the tag information can  be represented in an abstract embedding space, and also have the potential to capture the semantic relatedness among the tags.
Utilizing this ability in capturing the relatedness of POI tags, we have presented a word embedding based POI recommendation system.
We propose a novel tag-embedding based recommender system that models users and the potential places of interest.
Empirical evaluation on TREC Contextual Suggestion Track datasets validates the significant superiority of the proposed methods in comparison to state-of-the-art models developed for the same purpose.
the point of interest will further improve efficiency.
We use discrete optimization to find the best parameters instead of heuristically selecting them to test the robustness of the proposed methods.
The variations in results are demonstrated on altering the optimization measure to obtain the parameters for modeling users.
We further argue that the proposed method can minimize the risk of the privacy breach of the users. 
A limitation of the proposed model is that the POIs to be recommended to the users should have tags associated with them. 
As part of future works, we plan on exploring techniques similar to transfer learning for the same tasks to conquer the limitation of places with missing tags.
An immediate extension of the work would be to 
extend the venue-context appropriateness information by utilizing crawled data from location based social networks such as ~\cite{cross} to improve the performance of the method by better representing the places.



\section*{Acknowledgement}
The authors would like to thank the anonymous reviewers for their valuable comments and suggestions which have helped improve the quality of the reported work. Special thanks to Mohammad Aliannejadi for sharing the result files for the baselines reported in this article. 

\bibliographystyle{elsarticle-num}
\bibliography{contextual-suggestions}

\end{document}